\documentstyle[graphicx,axodraw,12pt,epsfig,epsf]{article} 
\begin{document}

\baselineskip 24pt

\newcommand{\sheptitle}
{Supersymmetry and electroweak breaking  with large and small extra 
  dimensions}

\newcommand{\shepauthor}
{V. Di Clemente, S. F. King and D. A. J. Rayner}

\newcommand{\shepaddress}
{Department of Physics and Astronomy,
University of Southampton, Southampton, SO17 1BJ, U.K.}

\newcommand{\shepabstract}
{We consider the problem of supersymmetry and electroweak breaking 
in a 5d theory compactified on an $S^{1}/Z_{2}$ orbifold, where
the extra dimension may be large or small. We consider the
case of a supersymmetry breaking 4d brane located at one of the orbifold
fixed points with the Standard Model gauge 
sector, third family and Higgs
fields in the 5d bulk, and the first two families on a parallel 4d matter
brane located at the other fixed point.  
We compute the Kaluza-Klein mass spectrum in this 
theory using a matrix
technique which allows us to interpolate between large and 
small extra dimensions.  We also consider the problem of electroweak symmetry 
breaking in this theory and localize the Yukawa couplings on the 4d matter
brane spatially separated from the brane where supersymmetry is broken.
We calculate the 1-loop effective potential 
using a zeta-function regularization technique,
and find that the dominant top and stop contributions are separately finite.
Using this result we find consistent electroweak symmetry breaking
for a compactification scale \mbox{$ 1/R \approx 830$ GeV} and a
lightest Higgs boson mass $m_{h} \approx 170$ GeV.}

\begin{titlepage}
\begin{flushright}
hep-ph/0107290\\
SHEP 01-18
\end{flushright}
\vspace{0.5in}
\begin{center}
{\Large{\bf \sheptitle}}
\vspace{0.5in}
\bigskip \\ \shepauthor \\ \mbox{} \\ {\it \shepaddress} \\
\vspace{0.5in}
{\bf Abstract} \bigskip \end{center} \setcounter{page}{0}
\shepabstract
\end{titlepage}

\section{Introduction} 

The Standard Model (SM) of electroweak and strong interactions provides a
 description of the fundamental particles and forces present in Nature. It has
been rigorously tested at high-energy colliders with excellent agreement.
However there are many theoretical reasons to believe that the SM is
 not a complete description of Nature.  Recently there have also been
experimental signals of ``new physics'' beyond the SM such as neutrino 
oscillations~\cite{nuoscillations} and discrepancies in $g_{\mu}-2$ 
measurements~\cite{g-2}.

An outstanding candidate for new physics beyond the SM is supersymmetry 
(SUSY) which solves many theoretical problems in a natural way and has lead to
a SUSY extension of the SM called the Minimal Supersymmetric Standard Model
(MSSM).  SUSY is also attractive since it is a fundamental symmetry in string 
theories, which provide the only consistent method of 
combining gravity with strong and electroweak forces in a single unified 
theory.  Recently there has been considerable interest in low-energy 
superstring-inspired models with heterotic (and now type I) models leading the
way towards fully realistic models. An unresolved problem 
is to understand the mechanism responsible for SUSY breaking, whereby 
supersymmetric partners acquire a large mass beyond the
reach of current accelerators.  This has been an active area of research for 
many years - some of the leading candidates are gravity~\cite{gravity},
gauge~\cite{gauge}, anomaly~\cite{anomaly} and gaugino mediated SUSY
breaking~\cite{gaugino}. Our previous work considered the embedding of gaugino
mediation into a type I string model involving intersecting 
D-branes~\cite{rayner}.

Regardless of ones opinion about superstring-inspired models involving 
D-branes, extra-dimensional ``brane world'' scenarios have become an active 
area of research in their own right.  They provide a
novel environment for investigating familiar problems such as
electroweak symmetry breaking (EWSB).  This 
problem has been the focus of much recent work in models involving 
large extra dimensions $(R \sim TeV^{-1})$ or equivalently low string 
scales $M_{\ast}$~\cite{antoniadis, delgado,barbieri,arkani1,delgado2}. The
models~\cite{delgado,barbieri,arkani1,delgado2} share similar features 
such as starting from a 5d theory then
compactifying the extra dimension on an $S^{1}/Z_{2}$ orbifold (with
the exception of \cite{barbieri} that has $S^{1}/Z_{2} \times Z'_{2}$ 
instead). This leads to fixed points invariant under 
$Z_{2}$-parity, where 4d D-branes can be located. SUSY is broken 
in the bulk by Scherk-Schwarz (SS) compactification\footnote{Notice that 
ref.~\cite{arkani1} also consider a scenario where SUSY is broken on a 
hidden sector brane.} of the 
fifth dimension~\cite{antoniadis, ss1,ss2,ss3,ss4}.  This compactification results in an 
infinite tower of Kaluza-Klein (KK) excitations for bulk fields, but not for
fields localized on either 4d brane. The 
Yukawa interactions are localised at the orbifold fixed points.  

The models \cite{delgado,barbieri,arkani1,delgado2} differ in the
type of SUSY breaking, the choice of orbifold and the
location of the MSSM fields, and in the 
methods used to analyse the spectrum and electroweak symmetry breaking.
Table 
\ref{tab:models} illustrates the important differences between these models.
\begin{table}[h]
 \begin{center}
 \scalebox{0.9}{
  \begin{tabular}{|c||c|c|c|c|} \hline
   Model & \cite{delgado} & \cite{barbieri} & \cite{arkani1} 
         & \cite{delgado2} \\ \hline\hline 
   Bulk fields & G,H & G,H,S,D & G,S,D & G,S \\ \hline
   Brane fields & S,D &  & H & H,D \\ \hline
  SUSY breaking & SS & SS & SS + SUSY brane & SS \\ \hline
  Higgs mass & $m_{h} \leq 110$ GeV & $m_{h} \sim 128$ GeV & &
      $m_{h} \leq 150$ GeV \\ \hline
  Compactification scale & $R^{-1} \sim 1$ TeV & $R^{-1} \sim 350$ GeV &
      $R^{-1} \sim 1$ TeV & $4 \leq R^{-1} \leq 10-15$ TeV \\ \hline
 \end{tabular}
               }
  \caption {{\small Comparison between the various models
~\cite{delgado,barbieri,arkani1,delgado2} showing where the gauge (G),
Higgs (H), $SU(2)_{L}$ singlets (S) and $SU(2)_{L}$ doublet fields (D) live in
the extra dimension.  The mechanisms that breaks SUSY are either the
Scherk-Schwarz (SS) boundary conditions or a SUSY breaking brane.  The models
also make EWSB predictions for the lightest Higgs boson mass $m_{h}$ and the
extra dimensional compactification scale $1/R$.  In model~\cite{arkani1}, the
``Higgs mass correction'' at 1-loop and zero external momenta is calculated.  
However this is not the physical mass since it corresponds to the second
derivative of the effective potential at $\langle H \rangle =0$.}}
  \label{tab:models} 
 \end{center}
\end{table}  

In this paper we consider a 5d theory compactified
on an $S^{1}/Z_{2}$ orbifold.
SUSY is broken on a 4d ``source'' brane located at one of the fixed points.
The first two MSSM families live on another 4d ``matter'' 
brane located at the other fixed point, while the
third family, MSSM gauge sector and the Higgs fields live in the extra 
dimensional bulk and therefore acquire non-trivial soft parameters due to 
their direct coupling to the SUSY breaking brane. 
This set-up, which differs from all the other models in 
Table \ref{tab:models}, is motivated by the string-inspired model
in \cite{rayner}. Notice that the presence of
the third family in the bulk, particularly the top and stop, is 
phenomenologically desirable for its important contribution to EWSB where
the up-like Higgs mass-squared is driven negative by 1-loop radiative
corrections which trigger the spontaneous breakdown of
${\mbox SU(2)_{L} \otimes U(1)_{Y} \rightarrow U(1)_{EM}}$ via the Higgs 
mechanism. Within this set-up we shall  
calculate the mass spectra of bulk field KK-resonances using two standard
methods, then using a matrix method which enables us to interpolate between
large and small extra dimensions and compare the results.
We will also consider the problem of EWSB in this extra dimensional model
where the top/stop Yukawa couplings are localized on the matter brane.
We calculate the
1-loop effective potential using dimensional regularization to perform the
momentum integral and zeta-function regularization to sum over an infinite
tower of KK-modes~\cite{kubyshin,hawking}.
We find that the top contribution is separately finite due to a
cancellation between the top and its CP-mirror field.  The stop 
contribution is also separately finite due to a cancellation
between stop and its CP-mirror fields and after cancellation gives
a constant contribution (independent of the Higgs background field). 
Therefore we find that
EWSB is triggered only by the finite 1-loop top contribution alone
which, unlike the stop contribution, depends on the Higgs background field.
If we neglect the Higgs interaction with the SUSY breaking 
brane and take $\tan \beta \rightarrow \infty$, minimization of the effective
potential allows us to make a prediction for the 
compactification scale $1/R\approx 830\ {\rm GeV}$  
and the lightest Higgs boson mass $m_h\approx 170\ {\rm GeV}$
which is heavier than for the models 
~\cite{delgado,barbieri,arkani1,delgado2} in Table \ref{tab:models}. 

The layout of the remainder of the paper is as follows.  In section 
\ref{sec:ourmodel} we introduce our string-inspired 5d model and discuss the 
${\mathcal N}=2$ SUSY formalism and allocation of MSSM fields.  In section 
\ref{sec:spectra}, we calculate the KK-mode 
mass spectra in the absence of Yukawa couplings, for 
models with a large or small extra dimension using three different methods. 
Section \ref{sec:yukawas} considers the localization of the Yukawa 
couplings on the matter brane and we revisit the third family mass
spectra in the presence of Yukawa couplings.  Then in section \ref{sec:ewsb}
we calculate the effective potential and discuss EWSB in this model.  Section 
\ref{sec:conc} concludes the paper.

\section{Our model}   \label{sec:ourmodel}

\subsection{Outline}  \label{sec:outline}

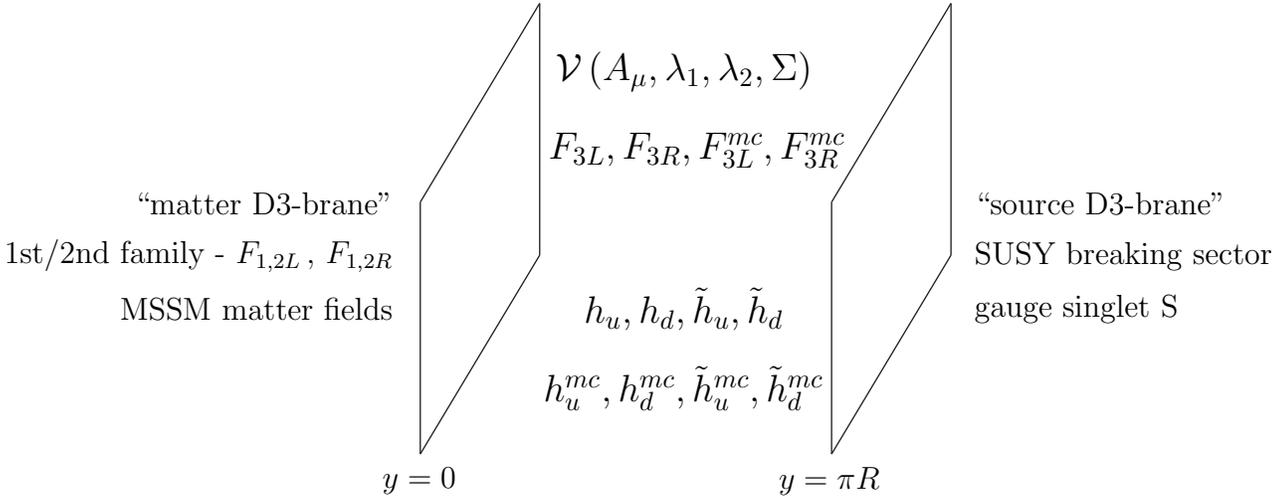
\begin{figure}[h]   
 \begin{center}  
  \begin{picture}(420,190)(0,0)
   \Line( 100, 110 )( 145, 185 )
   \Line( 100,  15 )( 145,  90 )
   \Line( 100,  15 )( 100, 110 )
   \Line( 145,  90 )( 145, 185 )
   \Text( 100,   5 )[c]{$y=0$}
   \Text(  90, 110 )[r]{``matter D3-brane''}
   \Text(  90,  90 )[r]{1st/2nd family - $F_{1,2L} \, , \, F_{1,2R}$}
   \Text(  90,  70 )[r]{MSSM matter fields}
   \Text(200,160)[]{{\large ${\mathcal V} \left( A_{\mu}, \lambda_{1}, 
      \lambda_{2}, \Sigma \right)$}}
   \Text(205,130)[]{{\large $F_{3L} , F_{3R} , F_{3L}^{m c}, 
      F_{3R}^{m c}$}}
   \Text(200,70)[]{{\large $h_{u} , h_{d} , \tilde{h}_{u} , \tilde{h}_{d}$}}
   \Text(200,40)[]{{\large $h_{u}^{m c} , h_{d}^{m c} , \tilde{h}_{u}^{m c} , 
      \tilde{h}_{d}^{m c}$}}
   \Line( 255, 110 )( 300, 185 ) 
   \Line( 255, 15 )( 300, 90 )
   \Line( 255, 15 )( 255, 110 ) 
   \Line( 300, 90 )( 300, 185 )
   \Text( 255, 5 )[c]{$y= \pi R$} 
   \Text( 310, 110 )[l]{``source D3-brane''} 
   \Text( 310, 90 )[l]{SUSY breaking sector} 
   \Text( 310, 70 )[l]{gauge singlet S} 
  \end{picture}
 \end{center} 
  \caption{{\small The model showing the parallel 3-branes spatially 
separated along the extra dimension y.  This extra dimension is compactified
on the orbifold $S^{1}/Z_{2}$ that leads to two fixed points at $y=0, \pi R$,
where the two D3-branes are located.  The first two chiral families live on 
the ``matter'' brane at $y=0$, while SUSY is broken by the F-term of a gauge 
singlet field S on the source brane at $y=\pi R$.  Following an explicit type 
I string construction, the third family, gauge fields, Higgses and Higgsinos
live in the extra dimensional bulk.  The bulk is required to be 
${\mathcal N}=2$ supersymmetric which requires the inclusion of MSSM
{\it ``mirror''} fields into the spectrum.  Yukawa couplings are localized on 
the matter brane at $y=0$.  The fields present in the model 
are summarised in Table \ref{tab:states}. }}    \label{fig:model}
\end{figure}

In this section we introduce our string-inspired model and discuss the 
location of the MSSM fields that arise from our string construction.  We will
also review the ${\mathcal N}=2$ formalism that is commonly used to describe 
supersymmetry in 5d.  The setup shown in 
Figure \ref{fig:model} is a simplification of our previous model, but with a 
{\it single} extra dimension compactified on an $S^{1}/Z_{2}$ 
orbifold~\cite{rayner}. 
\vspace*{-5mm}

From a
4d viewpoint, ${\mathcal N}=1$ SUSY in 5d is equivalent to ${\mathcal N}=2$ in
4d, since the Kaluza-Klein 
(KK) states can combine in pairs to form ${\mathcal N}=2$ states.
MSSM {\it mirror} fields also need to be added to the theory to respect the
${\mathcal N}=2$ SUSY and form hypermultiplets.
The $Z_{2}$-parity of the orbifold 
provides a classification of bulk (5d) fields into odd and even classes.  Odd 
fields vanish on the 4d branes at the fixed points, while even parity fields 
do not vanish
and can therefore couple to boundary fields.  Only even fields have $k=0$ 
KK-modes which can be associated with MSSM fields.  The mirror 
states are chosen to be odd and therefore do not appear in the MSSM spectra. 
Therefore
an ${\mathcal N}=1$ supersymmetric theory in 5d is equivalent to an
${\mathcal N}=2$ theory in 4d, where 5d bulk fields are equivalent to an
infinite tower of 4d KK resonances and ${\mathcal N}=2$ SUSY is
required for the KK modes to form Dirac masses in the bulk.  
The minimal supersymmetric multiplets in 5d are matter hypermultiplets
(chiral and  Higgs) and vector supermultiplets (gauge fields) constructed from
${\mathcal N}=1$ superfields and their ``mirror'' superfields. The definition 
of a mirror superfield is discussed in appendix \ref{app:2}.

The 5d vector supermultiplets ${\mathcal V}$ contain a five-dimensional gauge 
field $A_{M=\mu,5}$, a real scalar $\sigma$  and two Weyl fermions
$\lambda_{1,2}$ that all transform in the adjoint representation of the gauge 
group~\cite{ss4,sohnius}.  The 5d vector supermultiplet ${\mathcal V}$
can be decomposed into an ${\mathcal N}=1$
 vector supermultiplet (containing a gauge boson $A_{\mu}$ and a
gaugino $\lambda_{1}$) and an ${\mathcal N}=1$ chiral supermultiplet 
(containing a scalar $\Sigma \sim \sigma + i A_{5}$ and a
fermion $\lambda_{2}$). 

Similarly each 5d matter hypermultiplet can be decomposed into an
${\mathcal N}=1$ chiral supermultiplet and its CP-mirror
chiral supermultiplet.  For example, the up-like Higgs hypermultiplet 
${\mathcal H}_{u}$ contains the MSSM Higgs superfield 
$H_{u} \sim h_{u}, \tilde{h}_{u}$; and its CP-mirror 
$H_{u}^{mc} \sim h_{u}^{mc}, \tilde{h}_{u}^{mc}$.  Similarly for the other
matter hypermultiplets.  The model has twice
the particle content of the MSSM since bulk fields and their mirrors are both
needed to form ${\mathcal N}=2$ invariant states.  See appendix \ref{app:3} 
for a discussion of constructing fermion 4-component Dirac spinors from MSSM 
fields and the CP-conjugates of their mirrors.  The location of the fields 
present in our model are shown in Table \ref{tab:states}.
\begin{table}[h]
 \begin{center}
  \begin{tabular}{|c|c|c|} \hline
   States & Location & $Z_{2}$-parity \\ \hline\hline 
     \vrule width 0pt height 13pt
   $F_{iL} \sim Q_{iL},L_{iL}$ \hspace*{0.5cm} $(i=1,2)$ & $y=0$ &  \\ \hline
     \vrule width 0pt height 13pt
   $F_{iR} \sim U_{iR},D_{iR},E_{iR},N_{iR}$ \hspace*{0.5cm} 
     $(i=1,2)$ & $y=0$ &  \\ \hline
     \vrule width 0pt height 13pt
   $F_{3L} \sim Q_{3L},L_{3L}$ & bulk  & even \\ \hline
     \vrule width 0pt height 13pt
   $F_{3R} \sim U_{3R},D_{3R},E_{3R},N_{3R}$ & bulk & even \\ \hline
    \vrule width 0pt height 13pt
   $F_{3L}^{mc} \sim Q_{3L}^{mc},L_{3L}^{mc}$ & bulk  & odd \\ \hline
     \vrule width 0pt height 13pt
   $F_{3R}^{mc} \sim U_{3R}^{mc},D_{3R}^{mc},E_{3R}^{mc},N_{3R}^{mc}$ 
     & bulk & odd \\ \hline
     \vrule width 0pt height 13pt
   $V \sim A_{\mu},\lambda_{1}$ & bulk & even \\ \hline
     \vrule width 0pt height 13pt
   $\Sigma \sim \sigma+iA_{5},\lambda_{2}$ & bulk & odd \\ \hline
     \vrule width 0pt height 13pt
   $H_{u} \sim h_{u},\tilde{h}_{u}$ \hspace*{0.5cm} 
    $H_{d} \sim h_{d},\tilde{h}_{d}$ & bulk & even \\ \hline 
     \vrule width 0pt height 13pt
   $H_{u}^{mc} \sim h_{u}^{mc},\tilde{h}_{u}^{mc}$ \hspace*{0.5cm} 
    $H_{d}^{mc} \sim h_{d}^{mc},\tilde{h}_{d}^{mc}$ & bulk & odd \\ \hline 
     \vrule width 0pt height 13pt
   $S$ & $y= \pi R$ & \\ \hline 
 \end{tabular}
  \caption {{\small The location of the states present in our model.  Bulk 
fields are also classified by their transformation with respect to 
$Z_{2}$-parity.  Notice that the superfields Q, U, D, L, E, N implicitly 
include the scalar and fermion components, e.g. 
$Q_{iL} \sim \tilde{q}_{iL},q_{iL}$. }}
  \label{tab:states} 
 \end{center}
\end{table}  

There are two types of field present in the model: boundary fields that are
localized on either 4d brane, and bulk fields that feel the extra dimension
between the parallel 3-branes.  The compactification of this dimension on the
$S^{1}/ Z_{2}$ orbifold leads to a classification of the bulk superfields 
($\xi$) into odd ($\xi_{odd}$) and even ($\xi_{even}$) states, depending on 
their transformation under the $Z_{2}$ reflection $y \leftrightarrow -y$.
\newpage
\vspace*{-2cm}
\begin{eqnarray}
{\mathbf Even} \hspace*{5.3cm} {\mathbf Odd} \hspace*{2.2cm}\nonumber  \\
   \xi_{even} \left( x,-y \right) = \xi_{even} \left( x,y \right), 
    \hspace*{1.7cm}
     \xi_{odd} \left( x,-y \right) = - \xi_{odd} \left( x,y \right) 
        \hspace*{0.45cm}
\end{eqnarray}
The odd fields have KK expansions involving $\sin \left( ky/R \right)$ or 
$\sin \left( m_{k}y \right)$ where $k$ is the KK number and $m_{k}$ is the
$k^{th}$ KK-mode mass\footnote{Usually the KK modes have masses of the form
$m_{k} = k/R$.}.  They vanish at the fixed points, which means that odd
fields do not have zero modes which are associated with MSSM fields.
Whereas the even fields have $\cos \left( ky/R \right)$ or
$\cos \left( m_{k} y \right)$
expansions and therefore do not vanish at the orbifold fixed 
points\footnote{We can choose that the familiar MSSM fields are even with 
respect to the $Z_{2}$-symmetry, and so have massless zero modes 
before SUSY breaking.}.  These
$Z_{2}$-parity transformation properties are important when we come to couple 
bulk fields to boundary fields at either fixed point, for example in section
\ref{sec:yukawas} we localize the third family Yukawa couplings at $y=0$ using
a neat method involving an off-shell formulation of 
supersymmetry in 5d~\cite{mirabelli}.
 
\subsection{Lagrangian}

The 5d lagrangian can be split into an ${\mathcal N}=2$ invariant bulk 
term~\cite{sohnius} consisting of 5d bulk fields, and 4d ${\mathcal N}=1$
invariant brane terms localized on either 3-brane.  The 4d brane terms are 
formed from the boundary fields and the 4d even {\it projections} of the bulk 
fields on to the boundary branes.  There is a SUSY breaking term localized on
the source brane at $y=\pi R$.  The off-shell formalism of ${\mathcal N}=2$ 
SUSY in 5d is discussed in ref.~\cite{mirabelli}.
\begin{equation}
  {\mathcal L} = {\mathcal L}_{5} \left[ \,\, \xi 
   \left( x,y \right) \,\, \right] + \sum_{j} \delta
   \left( y-y_{j} \right) {\mathcal L}_{j} \left[ \,\, \xi \left(
   x,y_{j} \right), \eta_{j} \left( x \right) \,\, \right]
\end{equation}
   where j runs over the two branes at the orbifold fixed points,
     x are coordinates for the 4 non-compact dimensions, y is the coordinate
     for the extra compact spatial dimension, $\xi$ is a bulk field,
     and $\eta_{j}$ is a field localized on the $j^{th}$ brane.

The 5d lagrangian for vector ($A_{M}, \sigma, \lambda_{i}$) and matter 
hypermultiplets ($\Phi^{a}_{i}, \Psi_{a}$) given below 
~\cite{ss4,sohnius} includes the standard kinetic energy terms and
supersymmetric Yukawa interaction terms:
\begin{eqnarray}
 {\mathcal L}_{5} = Tr \frac{1}{g^{2}} \left\{ -\frac{1}{2} F_{MN}^{2}
+ \left| D_{M} \sigma \right|^{2} + i \bar{\lambda}_{i} \gamma^{M} D_{M} 
 \lambda^{i}
- \overline{\lambda}_{i} \left[ \sigma , \lambda^{i} \right] \right\} 
\hspace*{3cm} \nonumber \\
+ \left| D_{M} \Phi^{a}_{i} \right|^{2} 
+ i \overline{\Psi}_{a} \gamma^{M} D_{M} \Psi^{a}
- \left( i\sqrt{2} \Phi^{\dagger i}_{a} \overline{\lambda}_{i} \Psi^{a} + h.c.
 \right) - \overline{\Psi}_{a} \sigma \Psi^{a} 
\hspace*{1.5cm}  \label{eq:lagrangian} \\
- \Phi^{\dagger i}_{a} \sigma^{2} \Phi^{a}_{i}
- g^{2} \sum_{m,\alpha} \left[ \Phi^{\dagger i}_{a} 
 \left( \tau^{m} \right)^{j}_{i} T^{\alpha} \Phi^{a}_{j} \right]^{2}
\nonumber
\end{eqnarray}
where $a$ labels the bulk matter fields (including both Higgs
doublets and the third family superfields); $i,j=1,2$ are $SU(2)_{R}$
(R-parity) indices and $M,N=0-3,5$. $D_{M}$ is a covariant derivative and
$\tau^{m}$ are $SU(2)$ generators where m=1,2,3. 
 $\Phi^{a}_{i} (\Psi_{a})$ are the scalar 
(Dirac fermion) components of the Higgs and third family superfields.
   
Supersymmetry is broken by the F-term of a 4d gauge-singlet field S on the 
source brane at the fixed point $y= \pi R$ and mediated across the extra 
dimensional bulk by gauginos, third family scalars and Higgs fields as 
discussed in ref.~\cite{rayner}.
The source field couples directly to some of the even parity 5d bulk fields -
Higgses, gauginos and third family scalars - to form soft 
SUSY breaking terms localized at the $y= \pi R$ fixed point\footnote{Notice 
that if different gauge singlets on the source brane couple to different bulk 
fields then non-universal soft terms can arise.}.  The presence of
powers of the cutoff scale $M_{\ast}$ appear due to dimensional analysis and
the effective nature of the theory.

We have third family scalar masses and gaugino masses from the
following lagrangian:
\begin{eqnarray}
 \delta {\mathcal L}_{ \pi R}^{(1)} = \delta \left( y- \pi R \right) \left[
    -\int\!\! d^{4}\theta \,
   \frac{c_{F_{3L}}}{M_{\ast}^{3}} F_{3L}^{\dagger}
     F_{3L} S^{\dagger} S
    -\int\!\! d^{4}\theta \,
   \frac{c_{F_{3R}}}{M_{\ast}^{3}} F_{3R}^{\dagger}
     F_{3R} S^{\dagger} S  \right.  \hspace*{2cm} \label{eq:lagrange1}  \\
     \left.   + \int\!\! d^{2}\theta \, \frac{c_{w}}{16 g_{5}^{2} M_{\ast}^{2}} 
         S \, \, tr W^{\alpha} W_{\alpha} + h.c. \right] \nonumber
\end{eqnarray}
where $F_{3L}$ and $F_{3R}$ represent the third family superfields 
$Q_{3L} , L_{3L}, U_{3R} , D_{3R} , E_{3R} , N_{3R}$; $c_{F_{3L}}$ and
$c_{F_{3R}}$ are the coupling to the SUSY breaking field S;
 $g_{5}$ is the 5d
gauge coupling; and $W_{\alpha}$ is the 5d gauge field-strength superfield 
that contains the gaugino as its lowest component.

We can also generate soft Higgs masses, $B \mu$ and $\mu$-terms:
\begin{eqnarray}
 \delta {\mathcal L}_{ \pi R}^{(2)} = \delta \left( y- \pi R \right) \left[
    -\int d^{4}\theta \frac{1}{M_{\ast}^{3}} \left[
       c_{H_{u}} H_{u}^{\dagger} H_{u} + c_{H_{d}} H_{d}^{\dagger} H_{d}
          + c_{B \mu} H_{u} H_{d} + h.c.\right] S^{\dagger} S \right. 
        \nonumber \\
    \left. - \int d^{4}\theta \frac{c_{\mu}}{ M_{\ast}^{2}} 
        H_{u} H_{d} S^{\dagger} + h.c. \right] \label{eq:lagrange2}
\end{eqnarray} 
Notice that terms with even hypermultiplet fields replaced by their mirror
pairs are forbidden
by $Z_{2}$-parity as only even fields couple directly to the 3-brane 
boundaries at the orbifold fixed points.  However, the y-derivative of
an odd field is actually even with respect to $Z_{2}$-parity, so terms like
${\displaystyle \delta \left( y - \pi R \right) \int d^{4} \theta 
 \frac{1}{M_{\ast}^{5}} \partial_{y} Q_{3L}^{mc \dagger} \partial_{y}
 Q_{3L}^{mc} S^{\dagger} S}$ are allowed by the $Z_{2}$ symmetry, but are 
heavily suppressed by higher powers of the cutoff scale $M_{\ast}$, and can
therefore be neglected.

So far we have not specified where the Yukawa couplings arise in our 
model.  We adopt the standard approach of previous models and localize the 
Yukawas on either 4d brane since bulk Yukawa couplings explicitly break the
${\mathcal N}=2$ invariance\footnote{Notice that it {\it is} possible to 
construct higher dimensional operators in the bulk that respect the weaker 
constraint of $SU(2)_{R}$ (R-parity) invariance, but ${\mathcal N}=2$ SUSY is 
still explicitly broken in the bulk.}.  We will postpone this discussion until
section \ref{sec:yukawas}.

\section{Mass spectra - in the absence of Yukawa couplings}
 \label{sec:spectra}

In this section we will calculate the KK-mode mass spectra of some bulk fields
for large or small extra dimensions using two different standard methods.
First we will review the results for a small extra dimension where the 
non-zero KK-modes are effectively decoupled from the 
theory~\cite{gaugino,rayner}.  Then
we will use an equation-of-motion method developed in refs. 
~\cite{arkani1,chacko2,nomura} to find KK mass eigenvalues  using a KK
 expansion in terms of a mass 
eigenstate ansatz - $\cos (m_{k}y)$.  In section \ref{sec:spectra_meth3}
we will introduce a variation 
of a matrix method proposed in ref.~\cite{matrixmethod} that we feel is more 
powerful since it can (in principle) solve for large or small extra 
dimensions. 
We will find that the mass eigenvalues $m_{k}$ satisfy equations in terms of
the SUSY breaking parameters.  These relations can often be solved iteratively
by considering two different limits of strong and weak SUSY breaking.  
 We will show explicitly that the equation-of-motion method is
only applicable in either very strong or very weak SUSY breaking 
limits~\footnote{In particular, this method breaks down when we include 
Yukawa couplings and attempt to calculate the mass spectra.}.

\subsection{Mass spectra - small extra dimensions}   \label{sec:spectrasmall}

In the limit of a small extra dimension, we recover some results from the
original $\tilde{g}MSB$ model~\cite{gaugino} - but with the third family also 
in the bulk~\cite{rayner}.  Physically the compactification scale is very 
high,
of order the ultraviolet cutoff string scale, and so the KK-mode masses are
very heavy and effectively decouple from the theory.  Hence we only 
consider the ground state zero-modes (MSSM fields).

From our previous model~\cite{rayner} we have the following zero-mode mass 
predictions before EWSB.  They are expressed in terms of the cutoff 
scale $M_{\ast}$, the SUSY breaking parameter $F_{S}$ and a coupling 
parameter\footnote{See refs.~\cite{gaugino,rayner} for a discussion of
$\epsilon$, but essentially it represents the coupling strength of the 
theory.  $\epsilon \sim 1 $ for strong coupling.} $\epsilon$.
The first and second family scalar ($\phi_{1,2}$) soft masses are 
exponentially suppressed due to their displacement from the SUSY breaking 
sector.  Therefore these masses are negligible at the high scale.
\begin{equation}
 m_{\phi_{1,2}} = m_{\psi_{1,2}} = 0
\end{equation}
The third family scalars ($\phi_{3}$) couple directly to the source
brane to obtain a soft mass.
\begin{equation}
   m_{\phi_{3}}^{2} \sim \frac{1}{\epsilon l_{4}}
       \frac{ F_{S}^{2} }{M_{\ast}^{2}} , \hspace*{1cm}
   m_{\psi_{3}}=0
      \label{eq:chackomass}
\end{equation}
The MSSM gauginos ($\lambda$) are also coupled directly to the SUSY breaking 
and acquire soft masses, while the gauge bosons ($A_{\mu}$) do not.
\begin{equation}
  m_{\lambda} \sim \frac{1}{\epsilon l_{4}} \frac{ F_{S} }{M_{\ast}} ,
   \hspace*{1cm} m_{A_{\mu}} = 0
\end{equation}
The Higgs scalars ($h_{u}, h_{d}$) acquire soft masses, and their coupling 
also generates a mixing $B \mu$-term.
\begin{equation}
   B\mu , m_{h_{u}}^{2} , m_{h_{d}}^{2} \sim \frac{1}{\epsilon l_{4}} 
         \frac{ F_{S}^{2} }{M_{\ast}^{2}}   
\end{equation}
The $\mu$-problem is solved by the Giudice-Masiero 
mechanism~\cite{giudicemasiero} to give an {\it effective} soft mass to the
higgsinos ($\tilde{h}_{u}, \tilde{h}_{d}$).
\begin{equation}
    \mu \sim \frac{1}{\epsilon l_{4}} \frac{ F_{S} }{M_{\ast}} 
\end{equation}

These zero-mode predictions are approximate and arise from a naive dimensional
analysis.  In ref. ~\cite{rayner} we found that FCNC experimental data and the
desire for a phenomenologically valid ratio between gaugino and squark masses
leads to an allowed region of the coupling strength parameter 
$0.01 \leq \epsilon \leq 0.1$.

\subsection{Mass spectra - equation-of-motion method}
 \label{sec:spectralarge}

 We use a dynamical method, developed in 
refs.~\cite{arkani1,chacko2,nomura}, to find the mass eigenvalue $m_{k}$
 by proposing a mass eigenstate 
ansatz for the KK-mode expansion.  For example, in the case of weak SUSY
breaking, we would expect the KK-mode masses to be slightly perturbed away 
from the usual free-wave expansions - $\sin(ky/R)$ or $\cos(ky/R)$.  However
in the strong SUSY breaking limit, we expect the KK masses to be different 
- $\sin(m_{k}y)$ or $\cos(m_{k}y)$.  We obtain a set of coupled,
simultaneous differential equations in terms of the KK mass, that can usually
be solved iteratively.  This method works for both scalars
and fermions - where the requirement for ${\mathcal N}=2$ SUSY in the bulk 
couples odd and even parity fermion fields together.  This point is discussed 
in appendix \ref{app:3}.  Explicit details of the gaugino calculation are
given in appendix \ref{app:4}.

\subsubsection{Gauginos - $\lambda_{1,2}$}   \label{sec:m2gauginos}

The even ($\lambda_{1}$) and odd ($\lambda_{2}$) parity gauginos are combined 
together to form a 4-component Dirac spinor.  Following 
Eqn.\ref{eq:lagrange1}, we see that the even-parity gaugino 
$\lambda_{1}$ couples directly to the SUSY breaking sector at $y= \pi R$ and
acquires a localized soft mass.  However, the odd-parity ${\mathcal N}=2$ 
superpartner $\lambda_{2}$ is coupled to the even gaugino through the extra
dimensional kinetic term as shown in Eqn.\ref{eq:lagrangian} and appendix
\ref{app:3}.  

We can use the equation-of-motion method - discussed explicitly in appendix
\ref{app:4} for gauginos -
to obtain an expression relating the KK mass to the SUSY breaking F-term vev:
\begin{equation}
  \tan \left[ m_{\lambda , k} \pi R \right] = 
   \frac{c_{w} F_{S}}{4 M_{\ast}^{2}}  \label{eq:gauginos}
\end{equation}
which can be solved iteratively by treating $F_{S}$ as a small (large)
parameter for the weak (strong) SUSY breaking limit.

\subsubsection{Third family scalars - $\tilde{t}, \tilde{b}$}
  \label{sec:m23scalar}

Eqn.\ref{eq:lagrange1} shows that the third family scalars ($\tilde{t},
\tilde{b}$) live in the extra dimensional bulk and couple 
directly to the SUSY breaking brane to generate soft masses localized at
the $y=\pi R$ fixed point.  The mirror scalar partners are odd under the
$Z_{2}$-parity transformation and therefore do not couple to the SUSY
breaking sector to acquire soft masses.  Following the method of section 
\ref{sec:m2gauginos} and refs.~\cite{arkani1,nomura}, we find that the
mass eigenvalues (in the absence of Yukawa couplings) satisfy the following 
relation\footnote{Notice that with large $\tan \beta$, we are only interested
in the top/stop sector, particularly their contribution to the effective 
potential. Throughout the rest of the paper we will ignore the bottom 
contributions, except in this subsection where $m_{t,k}$ and $m_{b,k}$ 
coincide.}:
\begin{equation}
  m_{\tilde{t},k} \tan \left[ m_{\tilde{t},k} \pi R \right]
  = \frac{c_{\tilde{t}} F_{S}^{2}}{2 M_{\ast}^{3}}  \label{eq:mstop}
\end{equation}
%

\subsection{Mass spectra - matrix method} 
  \label{sec:spectra_meth3}

In this section we will discuss a matrix method that interpolates between
large 
{\it and} small extra dimensions.  It is a variant of a technique developed in
ref.~\cite{matrixmethod}.  We will give explicit examples of using the matrix
method to find the stop and gaugino KK-masses in the absence of Yukawa 
couplings.  The Higgs scalar mass spectrum is complicated by the presence
of the $B \mu$ and $\mu$ mixing parameters.  However, in the limit of a small 
$\mu$-term we find a spectrum similar to the stop KK-spectrum.  In the absence
of Yukawa couplings the top field has the usual KK mass spectra 
$m_{t,k} = k/R$ where ($k=0,1,2,\ldots$).

\subsubsection{Third family scalars - $\tilde{t} (\tilde{b})$}
  \label{sec:m33fscalars}

Now we will calculate the stop(sbottom) mass spectra using the matrix 
method\footnote{As discussed earlier, we are primarily interested in the
top/stop sector since they provide the dominant contributions to the 1-loop
effective potential due to the large size of the top/stop Yukawa coupling.}.
Using the following 5d KK-mode expansions:
\begin{eqnarray}
 \tilde{t}_{L}(x,y) &=& \frac{1}{\sqrt{2\pi R}} \tilde{t}_{L,0}(x) 
  + \sum_{k=1}^{\infty}\frac{1}{\sqrt{\pi R}} \cos \left( \frac{ky}{R}\right)
   \tilde{t}_{L,k}(x)   \label{fourierp} \\
 \tilde{t}_{L}^{mc}(x,y) &=& \sum_{k=1}^{\infty }\frac{1}{\sqrt{\pi R}} 
  \sin \left( \frac{ky}{R} \right) \tilde{t}_{L,k}^{mc}(x) 
    \label{fourieri}
\end{eqnarray}
and inserting this into Eqs.~\ref{eq:lagrangian} and \ref{eq:lagrange1},
we get the lagrangian mass term for the stop after integrating
over the extra dimension coordinate ($y$),
\begin{eqnarray}
 -{\mathcal L}_{4}^{mass} = \frac{1}{2R^2}\left[ \sum_{k=1}^{\infty} 
  k^{2} \left( \tilde{t}^{\ast}_{L,k} \tilde{t}_{L,k}
   + \tilde{t}^{mc \ast }_{L,k} \tilde{t}_{L,k}^{mc} \right)
 + \sum_{k,l=1}^{\infty} \frac{2\alpha}{\pi^2}(-1)^{k+l} 
  \tilde{t}_{L,k}^{\ast} \tilde{t}_{L,l} \right. \nonumber \\
 + \left. \sum_{k=1}^{\infty} \frac{2\alpha}{\sqrt{2} \pi^2}(-1)^{k} 
  \tilde{t}_{L,k}^{\ast} \tilde{t}_{L,0}
 + \frac{\alpha}{\pi^2} \tilde{t}_{L,0}^{\ast} \tilde{t}_{L,0} 
  + h.c. + \left( L \leftrightarrow R \right) \right]
    \label{lagrangian}  
\end{eqnarray}
where the dimensionless parameter is
\begin{equation}
 \alpha = c_{\tilde{t}} \pi \left( \frac{ F_{S}^{2}}{M_{\ast}^4} \right)
  M_{\ast}R        \label{eq:alpha}
\end{equation} 

The strong SUSY breaking
limit ($\alpha \rightarrow \infty$) corresponds to a large extra dimension 
since the soft mass $m_{soft}^{2} \sim 1/R^{2}$ and realistic phenomenology 
requires $1/R \approx {\mathcal O}(TeV)$.  Similarly in the weak limit 
($\alpha \rightarrow 0$), the soft mass $m_{0}^{2} \sim \alpha/R^{2}$ which 
requires $1/R \approx M_{P}$ which corresponds to a small extra dimension.
\begin{eqnarray}
  \alpha \rightarrow 0 \hspace*{1cm} & \Rightarrow & \hspace*{1cm}
 R \sim M_{P}^{-1} \\
  \alpha \rightarrow \infty \hspace*{1cm} & \Rightarrow & \hspace*{1cm}
 R \sim (TeV)^{-1}
\end{eqnarray}


Notice that the mixing between different KK-modes in the lagrangian of
Eqn.\ref{lagrangian} arise from the localization of the SUSY breaking on 
the brane at $y=\pi R$, and therefore the inclusion of a delta function in the
lagrangian\footnote{The presence of the brane at $y=\pi R$ explictly breaks 
the translational invariance along the extra dimension, and so the fifth 
dimensional momentum - and therefore KK number - is no longer conserved.}. 
The mass matrix is symmetric and may be written in the basis
$(\tilde{t}_{L,0} \,\, \tilde{t}_{L,1} \,\, \tilde{t}_{L,2} 
 \ldots )^{T}$ as,
\begin{equation}
{\cal M}^2 = \frac{1}{2R^2} \left(\matrix{
\alpha/\pi^2	                 &-\frac{2}{\sqrt{2}}\alpha/\pi^2     &\frac{2}{\sqrt{2}}\alpha/\pi^2	&-\frac{2}{\sqrt{2}}\alpha/\pi^2	
&\ldots\cr
-\frac{2}{\sqrt{2}}\alpha/\pi^2  & 1^2+2\alpha/\pi^2		      &-2\alpha/\pi^2			& 2\alpha/\pi^2		&\ldots\cr
\frac{2}{\sqrt{2}} \alpha/\pi^2	 & -2\alpha/\pi^2		      &2^2+2\alpha/\pi^2		&-2\alpha/\pi^2		&\ldots\cr
-\frac{2}{\sqrt{2}}\alpha/\pi^2	 & 2\alpha/\pi^2		      &-2\alpha/\pi^2			&3^2+2\alpha/\pi^2	&\ldots\cr
\vdots		 &\vdots			&\vdots			&\vdots			&\ddots
\cr}\right )
\end{equation}

Suppose that $\lambda^2$ is the eigenvalue associated with the eigenvector
 ${\mathcal Q}= (Q_0 \hspace*{3mm} Q_1 \hspace*{3mm} Q_2 \dots )^T$. We obtain 
the following set of eigenvalues equations ${\mathcal M}^{2} {\mathcal Q}
= (\lambda^{2}/(2 R^{2})) {\mathcal Q}$ which yield
\begin{eqnarray}
\alpha Q_0 -\frac{2}{\sqrt{2}}\alpha S_o + \frac{2}{\sqrt{2}} \alpha S_e &=& \pi^2\lambda^2 Q_0 \quad\qquad\qquad n=0 \label{q0}\\
-\frac{2}{\sqrt{2}}\alpha Q_0 + 2\alpha S_o - 2\alpha S_e &=& \pi^2(\lambda^2 - n^2) Q_n \quad\quad n\in odd
\\
\frac{2}{\sqrt{2}} \alpha Q_0 - 2\alpha S_o + 2 \alpha S_e &=& \pi^2(\lambda^2- n^2) Q_n \quad\quad n\in even
\end{eqnarray}
where
\begin{equation}
 S_o = \sum_{n \in odd} Q_n  \hspace*{1.5cm}
  S_e = \sum_{n\in even} Q_n
\end{equation}
It is straightforward to see that
\begin{eqnarray}
 S_e &=& \frac{2}{\sqrt{2}} \lambda^2 Q_0 \sum_{n\in even} 
  \frac{1}{\lambda^2 - n^2} \\
 S_o &=& \frac{\alpha}{\pi^2}\left( -\frac{2}{\sqrt{2}} Q_0
  + 2 S_o -2 S_e \right) \sum_{n\in odd} \frac{1}{\lambda^2 - n^2}
\end{eqnarray}
Following some algebra and using Eqn.\ref{q0} we find the relation:
\begin{equation}
 \pi^2\lambda^2 = \alpha + 2\alpha \lambda^2 
  \sum_{n=1}^{\infty}\frac{1}{\lambda^2 - n^2}
\end{equation}
where the physical mass eigenvalue is $m_{\tilde{t},k} = \lambda/R$ and we use the 
identity
\begin{equation}
 \sum_{n=1}^{\infty}\frac{1}{\lambda^2 - n^2} = 
  \frac{-1+ \pi\lambda\cot(\pi\lambda)}{2\lambda^2}
\end{equation}
to obtain the transcendental equation for the even stop (and sbottom) 
KK-mode masses{\footnote{Remember that the odd parity mirror fields remain 
massless since they do not couple to the even stop fields.}.
\begin{equation}
 m_{\tilde{t},k} \tan \left[ m_{\tilde{t},k} \pi R \right]
   = \frac{\alpha}{\pi R}
  \label{stop_mass}
\end{equation}

We can solve Eqn.\ref{stop_mass} iteratively by considering 
the limits of strong(weak) SUSY breaking, where $\alpha$ is a large(small) 
parameter~\cite{arkani1,nomura}. In the strong SUSY breaking limit 
$\alpha \gg 1$ (or equivalently $\sqrt{F_s} \sim M_{\ast}$) and the
extra dimension is large ($R M_\ast \gg 1$). In this case, 
Eqn.\ref{stop_mass} yields a spectrum where the low-lying mass
eigenvalues are approximately
\begin{equation}
 m_{\tilde{t},k} \approx \left(k + \frac{1}{2}\right)\frac{1}{R} 
  \left(1-\frac{1}{\alpha} + {\cal O}\left(\frac{1}{\alpha^2}\right)\right) 
   \hspace*{1cm} (k=0,1,2,\dots)
     \label{strong}
\end{equation}
Neglecting terms of order $1/\alpha$ and higher, we see that 
each KK-mode mass is shifted
up by half a unit relative to the usual unperturbed KK mass $k/R$.
The same conclusion applies when $\sqrt{F_{S}} \gg M_{\ast}$ and
$M_{\ast}R\approx 1$.  However it seems quite unnatural to have the SUSY
breaking $F$-term much larger than the cutoff of
the theory $M_{\ast}$. Physically, this implies that the SUSY breaking brane
acts as an impenetrable wall that makes the masses insensitive to the precise
values of $c_{\tilde{t}}$, $F_{S}^{2}$ and $M_{\ast}$, with the dependence
only arising in small higher-order corrections.  This limit can only be
phenomenologically viable if the compactification scale 
$1/R= {\mathcal O}(\mbox{TeV})$ i.e. if the extra dimension is {\it large}.

Compare the stop mass eigenvalues found using the equation-of-motion 
method (section \ref{sec:m23scalar}) and the matrix method (section 
\ref{sec:m33fscalars}) in the limit of strong SUSY breaking. 
We see that the resulting relations for the masses in
terms of SUSY breaking parameters - shown in Eqns.\ref{eq:mstop} and 
\ref{stop_mass} respectively - differ by a factor of $1/2$.  This difference
arises since the equation-of-motion method uses a KK-mode expansion where the
y-dependence of the input wavefunction is $\tilde{t}(x,y) \sim \sum_k f_{k}(y)
 \tilde{t}_{k}(x)$, where $f_{k}(y) \sim \cos(m_{\tilde{t},k} y)$ 
and $m_{\tilde{t},k}$ is the 
solution of Eqn.\ref{eq:mstop}.  However the y-dependence
of the wavefunction input in to the matrix method is $m_{\tilde{t},k}=k/R$
which explicitly generates a mass matrix which can be diagonalized to find
the physical mass eigenvalues.  The difference between these wavefunction 
profiles introduces some corrections into the mass term~\cite{nomura}. For 
example, consider the strong SUSY breaking limit where 
\begin{equation}
 \int_0^{\pi R} f_k(y) f_l(y) \sim \frac{R\pi}{2}\left(1+\frac{2}{\alpha}
  \right)\delta_{kl} + {\cal O}\left(\frac{1}{\alpha^2}\right)
\end{equation}
After integrating out the fifth dimension, the mass correction is given by
\begin{eqnarray}
 m_{\tilde{t},k}^{2} \sim \left(1+\frac{1}{2}\right)^2\frac{1}{R^2} 
  \left[ 1-\frac{2}{\alpha} + {\mathcal O}\left(\frac{1}{\alpha^2}\right)
   \right]
\end{eqnarray}
which agrees with the mass corrections found using the more general matrix
method (Eqn.\ref{strong}). 

\begin{figure*}[htb]
 \begin{center}
{\mbox{\epsfig{file=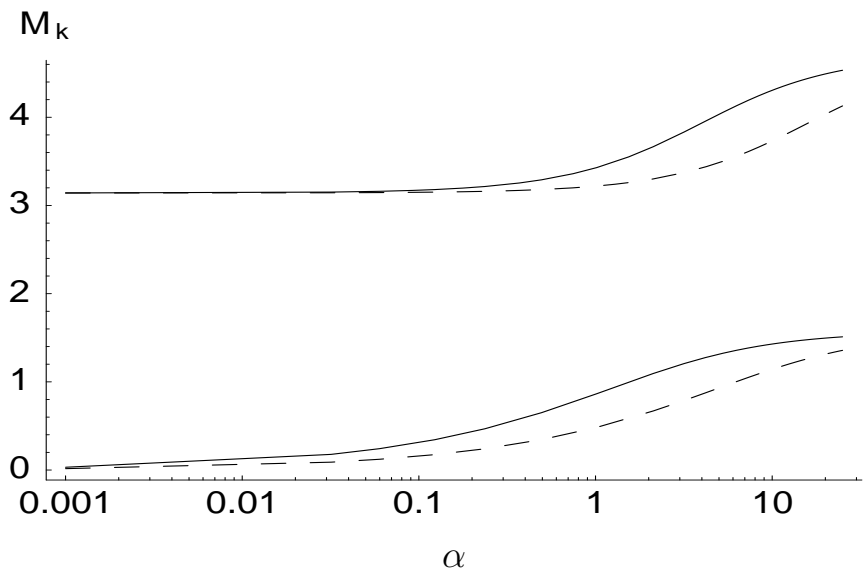,height=10.0cm,width=12.0cm}}}
\vskip-15mm
{\large $\mathbf \alpha$}
\caption[allfig]{Comparison of dimensionless stop mass parameter 
$M_{k}=m_{\tilde{t},k} \pi R$ against SUSY breaking parameter $\alpha$, where
$m_{\tilde{t},k}$ is the corresponding KK-mode mass.  Only the
first two KK-mode masses are shown - $k=0$ (lower) and $k=1$ (upper).  The 
continuous-line is the solution of 
Eqn.\ref{stop_mass} and the dashed-line arises from Eqn.\ref{eq:mstop}.}
   \label{figura1}
 \end{center}
\end{figure*}

However for arbitrary values of $\alpha$ the relationship between the two 
methods is more complicated. In general, the equation-of-motion method leads 
to a non-diagonal propagator in 4d.  In figure \ref{figura1} we have shown
the numerical difference between the two methods - considering the first two 
KK-excitations $k=0$ (lower) and $k=1$ (upper). As expected the two methods
converge in the limits $\alpha\to\infty$ and $\alpha\to 0$.  However the
region $0.1\leq\alpha\leq 10$ highlights a clear discrepancy, which becomes
more apparent for higher KK-excitations.  We can also conclude from 
$\alpha \sim 1$ that the extra dimension should have a compactification scale
$1/R \sim{\mathcal O}(TeV)$ to give phenomenologically reasonable zero-mode
soft masses\footnote{Remember that we associate the bulk field zero-modes with
the MSSM fields.}

In the weak supersymmetry limit ($\alpha \ll 1$) when $F_s \ll M_{\ast}$ and 
the fifth dimension length is small ($RM_{\ast} \sim 1$), we can utilize
effective field theory techniques to find the mass of the lightest KK-mode 
$(k=0)$ by decoupling higher(heavier) and lower(lighter) KK-modes. As a first
approximation we suppose that interactions between the heavier modes conserve
extra dimensional momentum (hence KK-number). 

Therefore, the lagrangian of
Eqn.\ref{lagrangian} can be rewritten as\footnote{This approximation is 
valid since the mixing between non-zero KK-modes do not affect the mass of
the zero mode.}: 
\begin{eqnarray}
 -{\mathcal L}_{4}^{mass} = \frac{\alpha}{2 \pi^{2} R^{2}} 
  \tilde{t}_{L,0}^{\ast} \tilde{t}_{L,0} 
 + \frac{1}{2R^2} \sum_{k=1}^{\infty} \left[ \left( k^{2} +
  \frac{2 \alpha}{\pi^{2}} \right) \tilde{t}^{\ast}_{L,k} \tilde{t}_{L,k}
 + \frac{2\alpha}{\sqrt{2} \pi^2}(-1)^{k} \left( \tilde{t}_{L,k}^{\ast}
  \tilde{t}_{L,0} + h.c. \right) \right]  \label{lagrangian1}  
\end{eqnarray}
where the dimensionless parameter $\alpha$ is given in Eqn.\ref{eq:alpha}

Integrating out the $\tilde{t}_{L,k}$ modes we obtain an expression for the
zero-mode mass-squared:
\begin{eqnarray}
 m_{k=0}^{2} &=& \frac{\alpha}{\pi^2 R^2}\left( 1 - 
  2\alpha\sum_{k=1}^{\infty}\frac{1}{k^2\pi^2 + 2\alpha}\right) \nonumber \\
 &=& \frac{3}{2}\frac{\alpha}{\pi^2 R^2}\left(1-
  \frac{\sqrt{2\alpha}}{3}\coth(\sqrt{2\alpha}) \right)  \label{mass0}
\end{eqnarray}
Expanding around $\alpha=0$ the zero-mode mass takes the compact form:
\begin{equation}
 m_{k=0}^{2} = \frac{\alpha}{\pi^2 R^2}\left(1- \frac{\alpha}{3} + 
  {\mathcal O}(\alpha^2)\right)   \label{mass00}
\end{equation}
Notice that the linear term in $\alpha$ comes directly from the lagrangian of
Eqn.\ref{lagrangian1} as expected.  The heavier modes now only contribute 
to the zero-mode mass at ${\cal O}(\alpha^2)$.

The weak SUSY breaking limit is very interesting because SUSY is broken at 
the high scale and the compactification scale ($1/R$) is also very large.  
This leads to a low-energy effective theory where SUSY is broken softly at a
much lower scale\footnote{Recently the same result has been observed in a 
similar model where SUSY is broken through the Scherk-Schwarz
mechanism~\cite{barbieri2} and the $\alpha$ parameter arises from a twisting
of the fields due to the requirement of translational invariance of the extra
dimension.} $\sim \alpha/R$.

We can also make a connection with the method of section 
\ref{sec:spectrasmall} if $M_{\ast}R \approx 1$ and $F_{S} \ll M_{\ast}$ 
($\alpha\ll 1$).  We find that the linear term in $\alpha$ in 
Eqn.\ref{mass00} gives the following zero-mode mass:
\begin{equation}
 m_{k=0}^{2} = \frac{c_{\tilde{t}} F_{S}^2}{\pi M_{\ast}^2}
   \label{zero-mode}
\end{equation}
Suppose that we set $c_{\tilde{t}}/\pi \approx 1/(\epsilon l_4)$.  We recover
the same relation for the zero-mode mass - Eqn.\ref{eq:chackomass} - found 
using the method for small extra dimensions in section \ref{sec:spectrasmall}.
In the small extra dimension scenario, the theory cutoff is associated with
the Planck scale $M_{\ast} \approx 1/R = 10^{19}GeV$.  This result implies 
that for
a realistic mass spectra ($m_0 = {\cal O}(\mbox{TeV})$), the SUSY breaking
$F$-term must be at an intermediate scale, $\sqrt{F_{S}} = 10^{11}GeV$. 
\begin{figure*}[htb]
 \begin{center}
{\mbox{\epsfig{file=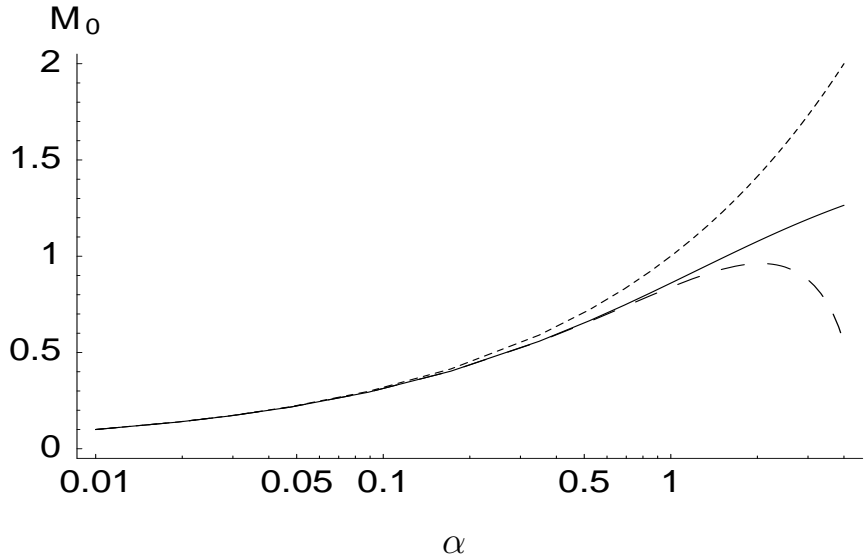,height=10.0cm,width=12.0cm}}}
\vskip-15mm
{\large $\mathbf \alpha$}
\caption[allfig]{Comparison of dimensionless stop mass parameter 
$M_{0}=m_{\tilde{t},0} \pi R$ against SUSY breaking parameter $\alpha$, where
$m_{\tilde{t},0}$ is the corresponding $k=0$ KK-mode mass.  The 
continuous-line is
the exact zero-mode solution of Eqn.\ref{stop_mass} and the dashed(dotted) 
lines are approximate solutions from Eqns.\ref{mass0} and \ref{mass00} 
respectively.  Notice that the dotted line only includes terms linear in 
$\alpha$ from Eqn.\ref{mass00}.}
   \label{figura2}
 \end{center}
\end{figure*}

We find the non-zero KK-mode masses by solving 
Eqn.\ref{stop_mass} in the limit $\alpha\ll 1$ 
\begin{equation}
 m_{\tilde{t},k} = \frac{k}{R}\left( 1 + \frac{\alpha}{k^2} 
  + {\mathcal O}(\alpha^2)\right) \hspace*{1cm} (k = 1,2,3,\dots)
\end{equation}
We see that for large KK-mode number, the spectra is approximately 
$m_{\tilde{t},k} \approx k/R$ which is the result we would expect since in 
this limit the SUSY breaking brane is completely ``transparent'' to the 
heavier KK-modes.

In figure \ref{figura2} we plot the zero-mode masses found numerically from
matrix method relations against the SUSY breaking parameter $\alpha$.  
Consider the zero-mode solutions of Eqn.\ref{stop_mass} (continuous-line),
the low energy spectra of Eqn.\ref{mass0} (dashed-line) and the mass found
by neglecting terms quadratic in $\alpha$ from Eqn.\ref{mass00} 
(dotted-line).  The deviation between the three zero-mode masses occurs for 
$\alpha\sim 1$, which implies that inside this range the decoupling theorem 
no longer works. A physical interpretation for $\alpha\geq 1$ is that the 
infinite tower of KK-modes {\it correlate} and behave like a single particle
and the decoupling of the lighter modes is non-trivial.  In this case, it is 
necessary to consider the interactions between all of the KK-modes.

\subsubsection{Gauginos  - $\lambda_{1,2}$} \label{sec:m3gauginos}

Following the discussion in section \ref{sec:m2gauginos} and appendix
\ref{app:4}, we know that the even parity
MSSM gaugino $\lambda_{1}$ and its odd ${\mathcal N}=2$ mirror $\lambda_{2}$ 
are coupled together in the 5d lagrangian.  The lagrangian mass terms are
\begin{equation}
 {\mathcal L}_{5}^{mass} = -\lambda_2\partial_5\lambda_1
  + \lambda_1\partial_5\lambda_2 - \frac{c_{w} F_{S}}{2 M_{\ast}^2}
   \delta(y-\pi R) \lambda_1\lambda_1 + h.c + \ldots
\end{equation}
Notice that only the even parity gaugino $\lambda_1$ couples to the 
supersymmetry breaking directly.

After integrating out the fifth dimension ($y$) and using the following
KK-mode expansions: 
\begin{eqnarray}
 \lambda_1(x,y) &=& \frac{1}{\sqrt{\pi R}}\lambda_0(x) + 
  \sqrt{\frac{2}{\pi R}}\sum_{k=1}^\infty 
   \cos\left(\frac{ky}{R}\right) \lambda_{1,k}(x) \\
 \lambda_2(x,y) &=& \sqrt{\frac{2}{\pi R}}\sum_{k=1}^\infty 
  \sin\left(\frac{ky}{R}\right) \lambda_{2,k}(x)  \label{eq:gauginoexp}
\end{eqnarray}
we can rewrite the 4d lagrangian associated with the gaugino mass term:
\begin{equation}
 -{\mathcal L}_{4}^{mass} = \frac{\beta}{\pi R}\lambda_{0} \lambda_{0} 
  + \frac{1}{R} \sum_{k,n=1}^{\infty} \left[ \left( 
   \sqrt{2}(-1)^k\frac{\beta}{\pi}\lambda_{0} \lambda_{1,k} 
    + 2(-1)^{k+n}\frac{\beta}{\pi}\lambda_{1,k}\lambda_{1,n} 
     - 2k \lambda_{1,k}\lambda_{2,k}\right) + h.c \right]
  \label{gaugino}
\end{equation}
where $\lambda_{0}=\lambda_{1,0}$ and $\beta= c_{w} F_{S}/(2M_{\ast}^2)$. 

Repeating the method discussed for stops in section \ref{sec:m33fscalars}, 
we can find the eigenvalues and eigenvectors of the mass matrix by choosing
the basis 
$(\lambda_{0} \hspace*{3mm} \lambda^{+}_{1} \hspace*{3mm} 
 \lambda^{-}_{1} \ldots \lambda^{+}_{k} \hspace*{3mm} \lambda^{-}_{k}
  \ldots)$ where
\begin{equation}
\lambda_{k}^{+} = \frac{\lambda_{1,k} + \lambda_{2,k}}{\sqrt{2}} 
 \qquad\quad \lambda_{k}^{-} = \frac{\lambda_{1,k} - \lambda_{2,k}}{\sqrt{2}} 
  \qquad (k\neq 0)
\end{equation}
In this basis, the gaugino mass matrix has the form:

\begin{equation}
{\mathcal M} = \frac{1}{\pi R} \left(\matrix{
\beta	 &-\beta     &-\beta	&\beta	  &\beta    &\ldots\cr
-\beta   & \beta -\pi&\beta     &-\beta	  &-\beta   &\ldots\cr	   
-\beta	 & \beta     &\beta+\pi &-\beta   &-\beta   &\ldots\cr
\beta	 & -\beta    &-\beta	&\beta-2\pi &\beta  &\ldots\cr
\beta	 & -\beta    &-\beta	&\beta	  &\beta+2\pi &\ldots\cr
\vdots   &\vdots     &\vdots	&\vdots	  &\vdots   &\ddots
\cr}\right )
  \label{gaugino_matrix}
\end{equation}

Suppose that the eigenvectors of ${\mathcal M}$ are found to be 
$(\Lambda_{0} \hspace*{3mm} \Lambda^{+}_{1} \hspace*{3mm} 
 \Lambda^{-}_{1} \ldots \Lambda^{+}_{k} \hspace*{3mm} \Lambda^{-}_{k}
  \ldots)$ with corresponding eigenvalue $\lambda$.  We obtain the following
eigenvalue equations for odd and even KK-modes:
\begin{equation}
 \hspace*{-2.5mm}
\beta\left(\Lambda_0 - S_o + S_e\right) = \lambda \Lambda_0 
  \hspace*{3.3cm} n=0 
\end{equation}
\begin{equation}
 \left. \begin{array}{c}
  \beta\left(-\Lambda_0 + S_o - S_e\right) = (\lambda+n\pi)\Lambda_n^+ \\
\vspace*{-2mm} \\
  \beta\left(-\Lambda_0 + S_o - S_e\right) = (\lambda-n\pi)\Lambda_n^- 
 \end{array}
  \right\} \hspace*{1cm} n \in odd
\end{equation}
\begin{equation}
 \left. \begin{array}{c}
  \beta\left(\Lambda_0 - S_o + S_e\right) = (\lambda+n\pi)\Lambda_n^+ \\
\vspace*{-2mm} \\
  \beta\left(\Lambda_0 - S_o + S_e\right) = (\lambda-n\pi)\Lambda_n^-
 \end{array}
  \right\} \hspace*{1.5cm} n \in even
\end{equation}
where
\begin{equation}
S_o = \sum_{n\in odd} \left(\Lambda_n^+ + \Lambda_n^-\right) \qquad\qquad 
S_e = \sum_{n\in even} \left(\Lambda_n^+ + \Lambda_n^-\right)
\end{equation}
After some algebra and using the identities
\begin{eqnarray}
 \sum_{n\in even} \frac{1}{\lambda^2 - n^2\pi^2} &=& \frac{-2 
  + \lambda \cot(\lambda/2)}{4\lambda^2} \\
  \sum_{n\in odd} \frac{1}{\lambda^2 - n^2\pi^2} &=& - 
   \frac{\tan(\lambda/2)}{4\lambda}
\end{eqnarray}
we find a relation for the mass eigenvalue $m_{\lambda,k} =
 \lambda/(\pi R)$ in terms of the SUSY breaking parameter $\beta$
\begin{equation}
 \tan(m_{\lambda, k}\pi R ) = \beta = \frac{c_{w} F_{S}}{2M_{\ast}^2} 
  \label{eq:gmass}
\end{equation}
which has the following solutions
\begin{equation}
 m_{\lambda,k} = \frac{k}{R} + \frac{1}{\pi R}\arctan(\beta)
  \hspace*{1cm} (k= 0, \pm 1 \pm 2, \dots)
\label{gauginomass}
\end{equation}
Solutions exist for both positive and negative $k$ 
where $k>0$ ($k<0$) is associated with
the eigenvector $\Lambda_k^+$ ($\Lambda_k^-$), however the
absolute value gives the physical mass. Hence, 
even though only the even gaugino $\lambda_{1}$
couples directly to the SUSY breaking sector, 
the two gaugino get mass due to mixing in the 5d kinetic  
term. Notice that this is completely different to the 
scalar sector where only the even parity
fields acquire masses.
\begin{figure*}[htb]
\begin{center}
{\mbox{\epsfig{file=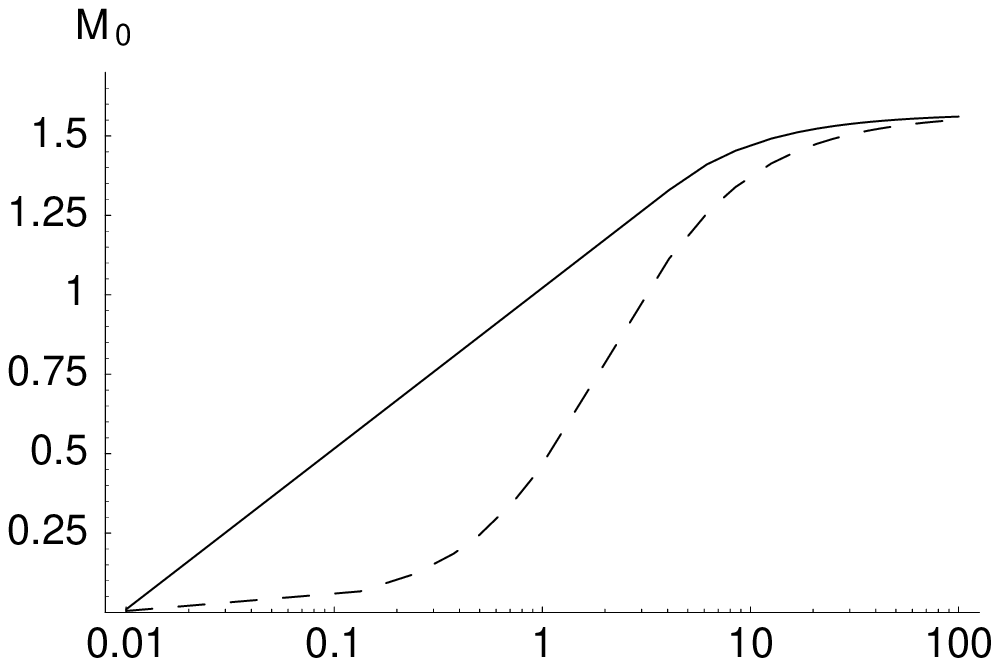,height=10.0cm,width=12.0cm}}}
\vskip-15mm
{\large $\mathbf \beta$}
\caption[allfig]{Comparison of dimensionless gaugino mass parameter 
$M_{0}=m_{\lambda,0} \, \pi R$ against SUSY breaking parameter $\beta$, where
$m_{\lambda,0}$ is the corresponding $k=0$ KK-mode mass.  The full line is
the zero-mode solution of Eqn.\ref{eq:gmass} using the matrix method, 
and the dashed line arises from the equation-of-motion method 
Eqn.\ref{eq:gauginos}.}
  \label{fig:gaugino_mass}
\end{center}
\end{figure*}

For example, in the case of weak SUSY breaking ($\beta \ll 1$), 
the masses are given by
\begin{equation}
 m_{\lambda,k} = \frac{k}{R} + \frac{\beta}{\pi R} - {\cal O}(\beta^3)
\end{equation}
which is the same mass spectrum that was obtained in ~\cite{daniel}
where supersymmetry was broken by an F-term defined in the bulk.
In the strong SUSY breaking limit ($\beta\gg 1$), we have
\begin{equation}
 m_{\lambda,k} = \left( k + \frac{1}{2} \right) \frac{1}{R} 
  - \frac{1}{\beta \pi R} + {\mathcal O} \left(\frac{1}{\beta^3} \right)
\end{equation}
Neglecting the $\beta$ 
contribution we get the same spectrum as in ~\cite{barbieri,arkani1},
where the compactification $S^1/(Z_2\times Z'_2)$ was used.

The different methods are related in a similar way to the stop analysis
in section \ref{sec:m33fscalars}. We will not repeat the full discussion 
here.  Figure \ref{fig:gaugino_mass} illustrates the differences in the
zero-mode mass found using the equation-of-motion method (dashed-line) and
the matrix method (continuous-line).  Notice that for the range 
$0.01 \leq \beta \leq 10$ there is a very large discrepancy between methods,
which suggests that the equation-of-motion method breaks down since the
propagator is non-diagonal in this region.

\section{Mass spectra - including Yukawa couplings}   \label{sec:yukawas}

In this section we will recalculate the stop mass spectra in the presence of 
Yukawa couplings using the matrix method of section \ref{sec:spectra_meth3}.
We adopt the conventional view of 
refs.~\cite{delgado,barbieri,arkani1,delgado2} and localize the Yukawa 
couplings on the {\it Yukawa brane} at $y=0$.  We will only consider the third
family Yukawa couplings since the the top/stop sector provides the dominant
radiative corrections to the 1-loop effective potential.  Notice that it is 
impossible to 
simultaneously maintain ${\mathcal N}=2$ SUSY in the bulk {\it and} have bulk 
Yukawa couplings since any higher-dimensional bulk operator that could 
generate bulk Yukawas explicitly breaks ${\mathcal N}=2$ SUSY.

We can immediately write down the 5d lagrangian term that generates top/stop 
Yukawa couplings on the Yukawa brane at $y=0$:
\begin{eqnarray}
 -{\mathcal L}_{5}^{Yuk} = \delta(y) \left[ 
   \frac{f_{t}}{M_{\ast}^{3/2}} \int\!\!d^{2}\theta \,
     Q_{3L} \cdot H_{u} U_{3R}^{c} + h.c. \right]   \label{eq:yukb1}
\end{eqnarray}
where $f_{t}=(2 \pi R M_{\ast})^{3/2} y_{t}$ and $f_{t} (y_{t})$ is the
5d(4d) Yukawa coupling.  Powers of the cutoff $M_{\ast}$ are required for
a lagrangian with the correct mass dimension\footnote{The lagrangian is 
required to have a total mass dimension of $5$.  Chiral superfields have a
mass dimension equal to that of the lowest component, the scalar field.  In
5d, the mass dimensions are: $[Q_{3L}]=[U_{3R}]=[H_{u}]=3/2$.  The 
$\theta$-integration and delta function have unit mass dimension, so the 
Yukawa couplings are dimensionless.}.

\subsection{Stop mass spectrum}  \label{sec:stopspec}

We begin by calculating the stop mass spectrum in the presence of Yukawa
couplings.
Following the method developed in ref.~\cite{mirabelli} for coupling bulk
fields to localized brane fields using an off-shell formulation of SUSY, we
can combine Eqns.\ref{eq:lagrangian} and \ref{eq:yukb1}.  We find 
the following expression for the auxiliary F-field $F_{Q_{3}}$
~\cite{mirabelli}:
\begin{eqnarray}
 F_{Q_{3}}^{\dagger} =  \delta(y) \, \frac{f_{t}}{M_{\ast}^{3/2}} \,
  \tilde{t}_{R} \, h_{u} - \partial_{5} \tilde{t}_{L}^{mc}
\end{eqnarray}
which can then be substituted back into the lagrangian Eq.\ref{eq:yukb1}
with the other F-terms.  However it is
important to notice the presence of a delta-function squared $\delta^{2}(y)$
that can be re-expressed as $\delta^{2}(y)=\delta(0) \delta(y)$. We recast
 $\delta(0)$ as
\begin{eqnarray}
 \delta(0) = \frac{1}{\pi R} \sum_{n=-\infty}^{\infty} 1
  = \frac{1}{\pi R} + \frac{2}{\pi R} \sum_{n=1}^{\infty} 1
  = \frac{2}{\pi R} \sum_{n=0}^{\infty} \left( \frac{1}{2}
   \right)^{\delta_{n0}} = \frac{2}{\pi R} {\mathcal D}
\label{D}
\end{eqnarray}
where ${\mathcal D}$ is an infinite quantity.

We can now integrate over the extra dimension to find the 4d lagrangian 
for the stop Yukawas in terms of KK modes\footnote{The Higgs field has the 
standard KK-mode expansion, but we assume that only the zero-mode acquires a 
non-zero vev that is identified with the corresponding MSSM vev.}, 
from Eq.\ref{eq:yukb1} we get:
\begin{eqnarray}
 -{\mathcal L}_{4}^{Yuk} = 8 y_{t}^{2} {\mathcal D} \sum_{k,l=1}^{\infty}
  \tilde{t}_{L,k}^{\ast} \tilde{t}_{L,l} h_{u,0}^{\ast} h_{u,0}
 + 4 y_{t}^{2} {\mathcal D} \tilde{t}_{L,0}^{\ast} \tilde{t}_{L,0} 
  h_{u,0}^{\ast} h_{u,0}
 + 4 \sqrt{2} y_{t}^{2} {\mathcal D} \sum_{k=1}^{\infty}
  \tilde{t}_{L,k}^{\ast} \tilde{t}_{L,0} h_{u,0}^{\ast} h_{u,0} 
   \nonumber \\
 - 2 y_{t} \sum_{k,l=1}^{\infty} \frac{k}{R} \,
  \tilde{t}_{R,k}^{mc \ast} \tilde{t}_{L,l} h_{u,0}
 - 2 y_{t} \sum_{k=1}^{\infty} \frac{k}{R} \,
  \tilde{t}_{R,k}^{mc \ast} \tilde{t}_{L,0} h_{u,0} + h.c.
 + \left( L \leftrightarrow R \right)  \label{eq:4dyuklag}  
\end{eqnarray}

Combine Eqs.\ref{lagrangian} and \ref{eq:4dyuklag} to rewrite the
4d stop mass terms including both soft SUSY breaking masses and Yukawa
contributions.
\begin{eqnarray}
 -{\mathcal L}_{4}^{mass} = \frac{1}{2R^2}\left[ \sum_{k=1}^{\infty} 
  k^{2} \left( \tilde{t}^{\ast}_{L,k} \tilde{t}_{L,k}
   + \tilde{t}^{mc \ast}_{R,k} \tilde{t}_{R,k}^{mc} \right)
 + \sum_{k,l=1}^{\infty} \left( \frac{2\alpha}{\pi^2}(-1)^{k+l} 
  + 16 m_{t}^{2} R^{2} {\mathcal D} \right) \tilde{t}_{L,k}^{\ast}
   \tilde{t}_{L,l} \right. \nonumber \\
 + \sum_{k=1}^{\infty} \left( \frac{2\alpha}{\sqrt{2} \pi^2}(-1)^{k} 
  + 8 \sqrt{2} m_{t}^{2} R^{2} {\mathcal D} \right) \tilde{t}_{L,k}^{\ast}
   \tilde{t}_{L,0}
 + \left( \frac{\alpha}{\pi^2}(-1)^{k} + 8 m_{t}^{2} R^{2} {\mathcal D} 
  \right) \tilde{t}_{L,0}^{\ast} \tilde{t}_{L,0} \\
 - \left. \sum_{k,l=1}^{\infty} 4k m_{t} R \left( \tilde{t}^{mc \ast}_{R,k} 
  \tilde{t}_{L,l} + \tilde{t}^{\ast}_{L,k} \tilde{t}_{R,l}^{mc} \right)
 - \sum_{k=1}^{\infty} 2 \sqrt{2} k m_{t} R \left( \tilde{t}^{mc \ast}_{R,k} 
\tilde{t}_{L,0} + \tilde{t}^{\ast}_{L,k} \tilde{t}_{R,0}^{mc} \right) \right] 
  + \left( L \leftrightarrow R \right) \nonumber   \label{eq:lagrangian2}
\end{eqnarray}
where we have replaced the Higgs field zero-mode by its vev $H_{c}$ so that
$m_{t}=y_{t} H_{c}$, and the SUSY breaking parameter is defined by
$\alpha = c_{\tilde{t}} 
\pi \left( F_{S}^{2}/M_{\ast}^{4} \right) M_{\ast}R$ as
in Eq.\ref{eq:alpha}.

We can now diagonalize the infinite mass matrix using the most convenient
basis:
\begin{equation}
 \left( \tilde{t}_{L,0} \hspace*{3mm} \tilde{t}_{L,k}  \hspace*{3mm}
  \tilde{t}_{R,l}^{mc} \right)^{T} \qquad\quad k,l = 1, \dots, \infty
\end{equation}
and the square mass matrix is:
\begin{equation}
{\cal M}^2 = \frac{1}{2R^2} \left(\begin{array}{ccc}
  \alpha/\pi^2 + 8 m_{t}^{2} R^{2} {\mathcal D}
& \sqrt{2} \alpha \,\tilde{I}/\pi^2 + 8 \sqrt{2} m_{t}^{2} R^{2} 
  {\mathcal D} \, I
& -2\sqrt{2} m_{t}R \, (I \cdot M) \\ \vspace*{-3mm}\\
  \sqrt{2} \alpha \, \tilde{I}^{T}/\pi^2 + 8 \sqrt{2} m_{t}^{2} R^{2} 
  {\mathcal D} \, I^{T}
& M^{2} + 2\alpha \, \tilde{J}/\pi^{2} 
  + 16 m_{t}^{2} R^{2} {\mathcal D} \, J
& -4 m_{t}R \, (J \cdot M) \\  \vspace*{-3mm}\\
 -2\sqrt{2} m_{t}R \, (M \cdot I^{T})
& -4 m_{t}R \, (M \cdot J)
& M^{2}
 \end{array}
\right)
\end{equation}
where $I=(1,1,1,1,\ldots)$, $\tilde{I}=(-1,1,-1,1,\ldots)$,
$M_{kl}=k \delta_{kl}$ and
\begin{equation}
 J=I^{T}I= \left(
  \begin{array}{cccc}
    1&1&1& \cdots \\
    1&1&1& \cdots \\
    1&1&1& \cdots \\
    \vdots & \vdots & \vdots & \ddots
  \end{array} \right)
\hspace*{1cm}
 \tilde{J}=\tilde{I}^{T} \tilde{I}= \left(
  \begin{array}{cccc}
    1&-1&1& \cdots \\
    -1&1&-1& \cdots \\
    1&-1&1& \cdots \\
    \vdots & \vdots & \vdots & \ddots
  \end{array} \right)
\end{equation}

Following a similar procedure as section \ref{sec:m33fscalars}, we can derive
an equation for the physical field-dependent mass-eigenvalue 
$m_{k}=\lambda/R$ in terms of the SUSY breaking parameters:
\begin{eqnarray}
 \lambda^{2}={\mathcal P}_{+} \pi \lambda \cot(\pi \lambda) 
  + \frac{\pi^{2}}{4} \left( {\mathcal P}_{+}^{2} + {\mathcal P}_{-}^{2}
   \right)
\label{lambdafinite}
\end{eqnarray}
where ${\mathcal P}_{\pm}= \pm \alpha/\pi^{2} + 4m_{t}^{2}R^{2} \pi \lambda
 \cot (\pi \lambda)$. This gives after some algebra the following 
transcendental equation~\footnote{Note that ${\mathcal D}$ defined in 
Eqn.\ref{D} has cancelled
to give a finite eigenvalue in Eqn.\ref{lambdafinite}}:
\begin{eqnarray}
 m_{\tilde{t},k} R \left[ \tan( \pi m_{\tilde{t},k} R )
  - \frac{4 m_{t}^{2} R^{2} \pi^{2}}{\tan( \pi m_{\tilde{t},k} R )} \right] 
   = \frac{\alpha}{\pi} \left( 1 + 4 m_{t}^{2} R^{2} \pi^{2} \right)
  \label{eq:tran1}
\end{eqnarray}
Once more we can solve this equation in the limit of large or small SUSY 
breaking.

\subsubsection{No SUSY breaking ($\alpha =0$)}

First consider the SUSY conserving limit ($\alpha =0$), Eqn.\ref{eq:tran1}
can be rewritten as
\begin{eqnarray}
 \tan^{2} ( \pi m_{\tilde{t},k} R ) = 4 m_{t}^{2} R^{2} \pi^{2}
\end{eqnarray} 
which has solutions
\begin{eqnarray}
 m_{\tilde{t},k}= \frac{k}{R} \pm \frac{1}{\pi R} 
 \arctan( 2 m_{t} \pi R)   \label{eq:tmass}
\end{eqnarray}
 
\subsubsection{Weak SUSY breaking ($\alpha \ll 1$)}

Now consider the weak SUSY breaking limit (small extra dimension) where 
Eqn.\ref{eq:tran1} can be solved in powers of the SUSY breaking parameter
$\alpha$.  The general result is very complicated, so we will only discuss the
zero-mode which has a mass eigenvalue with the following expansion:
\begin{eqnarray}
 m_{\tilde{t},k=0}^{2}= \frac{\alpha}{\pi^{2} R^{2}} + \frac{1}{\pi^{2} R^{2}}
  \left[ \arctan( 2 m_{t} \pi R) \right]^{2} 
   + \alpha^{2} f(2 m_{t} \pi R) 
    + \alpha^{3} g(2 m_{t} \pi R) + \ldots
  \label{eq:zerostop1}
\end{eqnarray}
where $f$ and $g$ are some functions of $(2 m_{t} \pi R)$.  We
can ignore the terms ${\mathcal O}(\alpha^{2})$ and higher when $\alpha \ll 1$
and we will see in section \ref{sec:topspec} that we recover a mass 
eigenvalue with the same field-dependence as the top.
However there is a discrepancy between the stop and top zero-mode masses when 
$\alpha \approx 1$ and the higher-order terms cannot be ignored necessarily.

\subsubsection{Strong SUSY breaking ($\alpha \longrightarrow \infty$)}
  \label{sec:sslargea}

Finally consider the strong SUSY breaking limit (large extra dimension) where
$\alpha \gg m_{t}$.  Eqn.\ref{eq:tran1} has a meaningful solution when
$m_{k}$ satisfies the following relations:
\begin{eqnarray}
 m_{\tilde{t},k} R \tan( \pi m_{\tilde{t},k} R )
  = \frac{\alpha}{\pi} \longrightarrow \infty  \\
{\mathrm or} \hspace*{5mm} 
  - m_{\tilde{t},k} R \cot( \pi m_{\tilde{t},k} R )
  = \frac{\alpha}{\pi} \longrightarrow \infty  
\end{eqnarray}
which yields the familiar eigenvalue of section \ref{sec:m23scalar} (if we 
ignore the small SUSY breaking parameter-dependent correction)
\begin{eqnarray}
  m_{\tilde{t},k}= \left( k + \frac{1}{2} \right) \frac{1}{R}
   \hspace*{1cm} \left( k= 0,\pm 1,\pm 2,\ldots \right)  
\label{eq:stopmassyuk} \\
{\mathrm and} \hspace*{5mm}
   m_{\tilde{t},k}= \left( 2k + 1 \right) \frac{1}{R}
   \hspace*{1cm} \left( k= 0,\pm 1,\pm 2,\ldots \right) 
\end{eqnarray}
Notice that the eigenvalue is independent of the Higgs vev which implies that 
in this limit  the stop contribution to the effective potential is
absorbed by the cosmological constant.  This implies that the strong SUSY 
breaking limit, in combination with a localized Yukawa coupling
brane,  {\it washes out} the field-dependence.  

\subsection{Top mass spectrum}  \label{sec:topspec}

Now we repeat the analysis for the top mass spectrum in the presence of the
Yukawa couplings.  Using the 5d lagrangian in appendix \ref{app:3} and 
Eqn.\ref{eq:yukb1}, we can write down the lagrangian terms that contribute
to the top mass:
\begin{eqnarray}
 {\mathcal L}^{mass}_{5} = t_{L}^{\dagger} \partial_{5} t_{L}^{mc} 
   - t_{L}^{mc \dagger} \partial_{5} t_{L} 
    + t_{R}^{\dagger} \partial_{5} t_{R}^{mc} 
     - t_{R}^{mc \dagger} \partial_{5} t_{R} 
      - \frac{f_{t} H_{c}}{M_{\ast}^{3/2}} \delta(y) 
       \left( t_{L}^{\dagger} t_{R} + t_{R}^{\dagger} t_{L} \right)  
 \label{eq:topmasslagr}
\end{eqnarray}
where $f_{t}= \left( 2 \pi R M_{\ast} \right)^{3/2} y_{t}$ and $f_{t}(y_{t})$
is the 5d(4d) Yukawa coupling; $H_{c}$ is the classical vev of the 
up-like Higgs field as before.  Using the familiar expansion in terms of
4d KK-modes, we obtain 
\begin{eqnarray}
 -{\mathcal L}_{4}^{mass} = - \sum_{k=1}^{\infty} \frac{k}{R} \left(  
  t^{\dagger}_{L,k} t_{L,k}^{mc} + t^{mc \dagger}_{R,k} t_{R,k} +h.c. \right)
 + 2m_{t} t_{L,0}^{\dagger} t_{R,0} \hspace*{3cm} \label{eq:topmasslagr2} \\
 + 2 \sqrt{2} m_{t} \sum_{k=1}^{\infty} \left( t_{L,0}^{\dagger} t_{R,k}
  + t_{L,k}^{\dagger} t_{R,0} + h.c. \right)
 + 4 m_{t} \sum_{k,l=1}^{\infty} \left( t_{L,k}^{\dagger} t_{R,k} + h.c.
  \right) \nonumber
\end{eqnarray}
This can be rewritten in block-diagonal matrix form:
\begin{equation}
 -{\mathcal L}_{4}^{mass} = \frac{1}{R} \sum_{k,l} \left(
  \begin{array}{ccc}
    t_{L,0}^{\dagger} & t_{L,k}^{\dagger} & t_{R,k}^{mc \dagger} 
  \end{array}  \right)
   \left( \begin{array}{ccc}
     2m_{t}R & 2\sqrt{2} m_{t} R \, I & 0 \\
     2\sqrt{2} m_{t} R \, I^{T} & 4m_{t}R \, (I^{T} I) & - M \\
     0 & - M & 0
   \end{array} \right)
    \left( \begin{array}{c}
      t_{R,0} \\
      t_{R,l} \\
      t_{L,l}^{mc}
    \end{array} \right)
     + \, h.c.
\end{equation}
where $I=(1,1,1,1,\ldots)$, $M_{kl}=k \delta_{kl}$ and $m_{t}=y_{t} H_{c}$.

We can derive a relation for the KK-mode mass eigenvalue in terms of the
classical Higgs vev:
\begin{eqnarray}
 \tan( \pi m_{t,k} R) = 2 m_{t} \pi R
\end{eqnarray}
which yield the mass eigenvalues
\begin{eqnarray}
 m_{t,k} = \frac{k}{R} + \frac{1}{\pi R} \arctan (2m_{t} \pi R )
  \hspace*{1cm} (k=0,\pm 1,\pm 2, \ldots)   \label{eq:topmassyuk}
\end{eqnarray}
This result is identical to the stop mass of Eqn.\ref{eq:tmass}
as expected since SUSY is not broken and it is  also consistent with
the result in ref.~\cite{barbieri}. Note that the zero mode top mass $m_{t,k=0}$ is different
from the usual 4D top mass.

\section{Electroweak Symmetry Breaking}   \label{sec:ewsb}

In this section, we will use the mass eigenvalues found in section
\ref{sec:yukawas} with Yukawa couplings on the brane, to calculate
the 1-loop effective potential.  We follow a method developed in 
ref.~\cite{kubyshin} and consider the top/stop contributions to a
summation over the infinite tower of KK-modes.  We will minimize the
effective potential to determine whether EWSB is possible in the limit of
$\tan \beta \rightarrow \infty$.  We also obtain a prediction for the 
lightest scalar Higgs mass.

We are interested in calculating the dominant radiative corrections to the 
1-loop effective potential arising from the top/stop sector\footnote{The
top/stop fields give the largest contribution to the 1-loop effective 
potential due to the large size of their Yukawa coupling relative to the 
other fields.}.  From Eqns.\ref{eq:stopmassyuk} and \ref{eq:topmassyuk}
we have the following field-dependent masses for the top and stop fields:
\begin{eqnarray}
  m_{t,k}(H_{c}) &=& \left| \pm \frac{k}{R} + \frac{1}{\pi R}
   \arctan(2 \,y_{t} H_{c} \pi R ) \right| \\
  m_{\tilde{t},k}^{2}(H_{c}) &=& \left( \pm k + \frac{1}{2}
   \right)^{2} \!\!\!\frac{1}{R^{2}} 
\end{eqnarray}

The 1-loop effective potential is given by the formula
\begin{eqnarray}
 V_{1-loop} = \frac{1}{2} Tr \sum_{k=-\infty}^{\infty}
 \int \!\! \frac{d^{4}p}{ \left( 2 \pi 
  \right)^{4}}  
   \ln \left[ \frac{ R^{2} \left( p^{2} 
   + m_{\tilde{t},k}^{2}(H_{c}) \right)}{R^{2} \left( p^{2} 
    + m_{t,k}^{2}(H_{c}) \right)} \right] 
 = V_{\tilde{t}}\, (H_{c}) + V_{t}\, (H_{c})   \label{eq:1loop}
\end{eqnarray}
where the trace is over all degrees of freedom.  Each top/stop (and mirror)
field has three colours and 
four degrees of freedom, i.e. a four-component fermion or two complex 
scalars.  The trace gives an overall factor of $N_{c}=12$.

Interchanging 
the summation over KK-modes with the integration over momenta, one
finds ~\cite{delgado,barbieri,arkani1}
\begin{eqnarray}
V_{\tilde{t}} &=& N_c \int \!\! \frac{d^{4}p}{ \left(2 \pi 
  \right)^{4}}\left( \pi R p + \ln \left(1 + e^{-2\pi R p}\right) \right) 
\label{Vstop} \\
 V_{t} &=& -\frac{N_c}{2} \int \!\! \frac{d^{4}p}{ \left(2 \pi 
  \right)^{4}}\left( 2 \pi R p + \ln \left(1- r e^{-2\pi R p}\right) + 
\ln \left(1- \frac{1}{r} e^{-2\pi R p}\right) \right) \label{Vtop}
\end{eqnarray}
where $r = \exp(2i \arctan(2 y_t H_c \pi R))$. When we calculate 
these integrals using a cutoff we see the ``infinite contribution'' 
which comes from the first terms $\pi R p$ of
Eqns.\ref{Vstop} and \ref{Vtop} cancels out in the final 
expression~\ref{eq:1loop}.
The remaining terms give a finite contribution
because they are exponentialy suppressed with respect to the momentum.  

Alternatively we can first perform the momentum
integration using standard dimensional regularization,
leading to a result which consists of an infinite
pole part proportional to $1/\epsilon$ plus a finite part.
Then we can argue that the pole term gives a zero
contribution once the KK sums are properly regulated
using zeta-function regularization.

There has recently been some criticism regarding how to calculate the 
effective potential~\cite{kkregcrit}.  The first part of the 
criticism states that for
each KK-mode, the integral in Eqn.\ref{eq:1loop} is ultraviolet-divergent,
and so it
is unclear whether it is possible to interchange the summation with the
momentum integral. The second part of the 
criticism is concerned that KK-modes 
{\it above} the theory cutoff can contribute to the effective potential. 
The zeta-function regularization approach successfully overcomes 
the first problem by using dimensional regularization to evaluate
the momentum integrals first, 
and then afterwards performing the infinite sum using zeta-function
regularization~\cite{kubyshin}. 
Since in this approach we integrate over all 4d momenta, it is natural
that we should also sum over all KK number, 
since the KK number is related to
momentum flowing along the direction of the extra dimension,
so in this approach
the second criticism does not apply either.
\footnote{In approaches where both a momentum cut-off and 
an appropriate symmetry-preserving KK cut-off are used, 
the heavier KK-mode contributions are found to be exponentially
suppressed~\cite{kkregcrit2}. }

Performing the integrals in Eq.\ref{eq:1loop} using dimensional 
regularization, 
the infinite part of the effective potential arising from the top contribution
can be expressed as ~\cite{kubyshin}:
\begin{eqnarray}
 V_{t}^{\infty} \, (H_{c}) = \frac{N_{c}}{32 \pi^{2} \epsilon} \left[
\left( m_{t,k=0} \right)^{4} + \sum_{k=1}^{\infty} \left( \frac{k}{R} +
m_{t,k=0} \right)^{4} + \sum_{k=1}^{\infty} \left( - \frac{k}{R} +
m_{t,k=0} \right)^{4} \right]  \label{eq:vinfty}
\end{eqnarray}
Then, using
zeta-function regularization ~\cite{kubyshin,hawking}
to perform the infinite summation, we find that
the three terms in Eqn.\ref{eq:vinfty} cancel each other, and
$V_{t}^{\infty} \, (H_{c})$ is exactly zero.  This cancellation occurs due
to the explicit ${\mathcal N}=2$ SUSY in the bulk that would otherwise not
occur for models where the higher KK-modes are decoupled or the
infinite sum is truncated. 
For a non-supersymmetric model where only the first two terms in 
Eqn.\ref{eq:vinfty} arise, the cancellation also would not occur.
Note that in this approach the top contribution to the effective
potential is therefore finite and the infinite contribution in 
Eq.\ref{Vtop} does not appear, because in dimensional regularization
this is defined to be zero. Similarly the stop contribution
also gives a finite contribution and the infinite contribution
in Eq.\ref{Vstop} also does not appear.

The finite top contribution is:
\begin{eqnarray}
 V_{t} \, (H_{c}) &=&  \frac{ 3 N_{c}}{64 \pi^{6} R^{4}}
  \sum_{n=1}^{\infty} \frac{\cos \left[ 2 \pi R n m_{t,k=0}(H_{c}) 
   \right]}{n^{5}}
     \label{eq:topcont}
\end{eqnarray}
which coincides with the finite part of Eqn.\ref{Vtop}.
\noindent Similarly, the stop contribution is a finite constant:
\begin{eqnarray}
 V_{\tilde{t}} \, (H_{c}) &=& - \frac{ 3 N_{c}}{64 \pi^{6} R^{4}}
  \sum_{n=1}^{\infty} \frac{ (-1)^{n}}{n^{5}}  \nonumber \\
 &=& \frac{45 N_{c} \zeta(5)}{1024 \, \pi^{6} R^{4}}   \label{eq:stopcont}
\end{eqnarray}
again this result is the same obtained in the finite part of Eqn.\ref{Vstop}.

Notice that in the case of small extra dimensions, the non-zero KK-modes are
decoupled from the zero-mode in an effective theory.  Hence, when we calculate
the infinite part of the effective potential $V_{t,\tilde{t}}^{\infty}(H_{c})$
we only have the first term ($k=0$) in Eqn.\ref{eq:vinfty}, and therefore
recover the familiar MSSM effective potential.

In general, the tree-level effective potential is given by:
\begin{eqnarray}
 V_{tree} = m^{2} H_{c}^{2} + \frac{g^{2} + g'^{2}}{8} H_{c}^{4}
  + \Lambda
\end{eqnarray}
where $g$ and $g'$ are the gauge couplings of $SU(2)_{L}$ and $U(1)_{Y}$
respectively; $\Lambda$ is the cosmological constant; and $m^{2}$ is a soft 
mass generated by the coupling of the Higgs field to the SUSY breaking brane
\begin{eqnarray}
  m^{2} = \frac{c_{H_{u}} \alpha}{2 c_{\tilde{t}} R^{2}}  \label{eq:higgssoft}
\end{eqnarray}
where $\alpha$ is given by Eqn.\ref{eq:alpha}.  
 $c_{H_{u}}, c_{\tilde{t}}$ are the coupling constants from 
Eqn.\ref{eq:lagrange2}.

In the strong SUSY breaking limit ($\alpha \gg 1$) and $m^{2} \gg 0$.
  This is a significant problem in order for EWSB
to be triggered via radiative corrections.  One possible solution is to 
introduce a hierarchy between $c_{H_{u}}$ and $c_{\tilde{t}}$, or 
alternatively assume that the Higgses and third family couple to different
gauge-singlets on the SUSY breaking brane (non-universality).  However, we 
will make a simplification and set $c_{H_{u}}=0$ in what follows.

The total effective potential is now
\begin{eqnarray}
 V_{eff}(H_{c}) = \frac{g^{2} + g'^{2}}{8} H_{c}^{4}
  + \tilde{\Lambda} + \frac{ 3 N_{c}}{64 \pi^{6} R^{4}}
  \sum_{n=1}^{\infty} \frac{\cos \left[ 2 \pi R n m_{t,k=0}(H_{c}) 
   \right]}{n^{5}}
    \label{eq:veff}
\end{eqnarray}
where we have absorbed the constant stop contribution into a redefinition of
the cosmological constant $\tilde{\Lambda}$.

Taking the second derivative at the origin gives:
\begin{eqnarray}
 \left. \frac{ d^{2} V_{eff} }{d H_{c}^{2}} \right|_{H_{c}=0} 
  = \frac{ -3 N_{c} y_{t}^{2} \zeta(3)}{4 \pi^{4} R^{2}}
     \label{eq:second}
\end{eqnarray}
Then, EWSB is triggered by radiative corrections at the compactification 
scale. The last equation is consistent with the result using Feynman 
diagrams to evaluate the
Higgs scalar two-point function at zero external momenta~\cite{arkani1}.  
However this mass is not identified with the physical Higgs boson mass.

Using the $\overline{{\mbox MS}}$ top mass $m_{t,0} = 166$ GeV, we can 
calculate the
compactification scale by imposing the following minimization conditions
around $v=175$ GeV:
\begin{eqnarray}
 \left. \frac{ d V_{eff} }{d H_{c}} \right|_{H_{c}=v} &=& 0 
   \label{eq:first} \\
 \left. \frac{ d^{2} V_{eff} }{d H_{c}^{2}} \right|_{H_{c}=v} 
   &=& m_{h}^{2}  \label{eq:veffmass}
\end{eqnarray}

Numerically we find:
\begin{equation}
 \frac{1}{R} \approx 830 \,\,{\mathrm GeV}
\end{equation} 
which is approximately $2.5$ times larger than the compactification scale 
calculated in
ref.~\cite{barbieri}.  The second-derivative of the effective potential at
the vacuum $(H_{c})$ yields a lightest Higgs scalar mass
\begin{equation}
  m_{h} \approx 170 \,\, {\mathrm GeV}
\end{equation}

We have seen that in the strong SUSY breaking limit, the stop KK mass spectra
is independent of the background Higgs field which gives a constant
contribution to the effective potential.  Electroweak symmetry breaking is
radiatively triggered by the top sector at 1-loop.  However these results
cannot be generalized to the Standard Model embedded in extra dimensions
since ${\mathcal N}=2$ SUSY is required for a well-defined theory.

These predictions apply for our simplified toy model where we have set
$\tan \beta \rightarrow \infty$ and neglected the coupling of the Higgs
fields to the SUSY breaking brane which would generate soft masses.  It will
be interesting to repeat our analysis in a more general two Higgs doublet
model with more realistic values of $\tan \beta$.

\section{Conclusions}  \label{sec:conc}

In conclusion we have
considered the problem of supersymmetry and electroweak breaking 
in a 5d theory compactified on an $S^{1}/Z_{2}$ orbifold, where
the extra dimension may be large or small. In our model there is
a supersymmetry breaking 4d brane located at one of the orbifold
fixed points with the Standard Model gauge 
sector, third family and Higgs
fields in the 5d bulk, and the first two families on a parallel 4d matter
brane located at the other fixed point.  
This set-up was motivated by a recent string-inspired
analysis \cite{rayner}, and as discussed in that model
will lead to a characteristic SUSY mass spectrum 
where the third family sparticles are heavier than
the second family sparticles, which only receive masses
through radiative corrections, thereby solving the
SUSY flavour changing neutral current problem.
We have computed the Kaluza-Klein mass spectrum in this 
theory using a matrix 
technique which allows us to interpolate between large and 
small extra dimensions, ranging from the GUT scale
down to a TeV. The matrix method was shown to lead to a more reliable
estimate of the mass spectrum especially in the parameter   
regions $\alpha , \beta \sim  0.1-10$ which may be relevant for
TeV scale extra dimensions.

Using the reliable matrix method we have also 
calculated the KK mass spectra including the important third family
Yukawa couplings on the matter brane.
Remarkably, in the strong SUSY breaking limit, 
background (Higgs) field dependence of the stop mass is
washed out and the only remaining field dependence is
in the top mass. Using these results we calculated the 
1-loop effective potential, including
the dominant top and stop corrections.
By performing the momentum integrations first, using standard
dimensional regularization, then regulating the KK sums
using zeta-function regularization, we obtained a finite result
for the 1-loop effective potential which is not subject to the
criticisms that have been made concerning previous approaches.
The resulting effective potential has a second derivative at the
origin which agrees with a recent Feynman diagram calculation,
which gives us some confidence in our effective potential.
By mininising our effective potential in the limit of large 
$\tan \beta$, and making the simple approximations that the
tree level Higgs mass and $\mu$ parameter are both zero,
enables us to predict the compactification scale 
\mbox{$ 1/R \sim 830$ GeV} and the lightest Higgs boson mass $m_{h} \approx
170$ GeV. Our Higgs mass prediction is significantly
higher than other models based on SS SUSY breaking.
Of course in the limit of small extra dimensions the model
reduces to the MSSM with usual radiative electroweak symmetry 
breaking and the standard Higgs mass bound applies.

\newpage
\begin{center}
{\bf \large Acknowledgements}
\end{center}
S.K., V.D.C. and D.R. would like to thank PPARC for a Senior Fellowship,
Research Associateship and a Studentship.  We would like to thank R.Barbieri,
A.Delgado, Y.Kubyshin and M.Quir\'os for useful discussions.

\appendix

\section{Appendix}
\subsection{5d Dirac matrices}  \label{app:1}

In this appendix we will review the Dirac matrices that appear in the 
fermion terms of the 5d Lagrangian.  We will use the notation that the indices
M, N run over 0,1,2,3,5; and $\mu$ runs over 0,1,2,3 as usual.  We use a 
timelike metric ${\displaystyle \eta_{MN}=diag \left(1, -1, -1, -1, -1
 \right) }$, and take the following basis for the 5d Dirac matrices:
\begin{equation}
 \gamma^{M}= \left[ \left(
   \begin{array}{cc} 
     0 & \sigma^{\mu} \\
     \overline{\sigma}^{\mu} & 0 \end{array} \right) ,
 \left(  \begin{array}{cc} 
     -i & 0 \\
     0 & i \end{array} \right) \right]
\end{equation}
where ${\displaystyle \sigma^{\mu}= \left( 1, \underline{{\bf \sigma}}
 \right) }$ and ${\displaystyle \overline{\sigma}^{\mu}= 
 \left( 1, -\underline{{\bf \sigma}} \right) }$.

\subsection{Mirror fields}  \label{app:2}

${\mathcal N}=2$ SUSY requires that the {\it mirror} partners of
${\mathcal N}=1$ MSSM superfields need to be included to construct
${\mathcal N}=2$ hypermultiplets.  We will remove any ambiguity by specifying
what we mean by a ``mirror'' partner.
Consider the left-handed quark MSSM doublet, $Q_{L}$,
as an explicit example.  Under the MSSM ``321'' gauge symmetry and Lorentz 
symmetry, $Q_{L}$ has the following quantum numbers respectively:
\begin{equation}
 Q_{L}: \hspace*{0.5cm} \left( \underline{3}, \underline{2}, \frac{1}{6}
   \right) \hspace*{1.5cm} \left( \frac{1}{2}, 0 \right)
\end{equation}
Now the mirror $Q_{L}^{m}$ has the opposite gauge quantum numbers, but still 
transforms like a left-handed field:
\begin{equation}
 Q_{L}^{m}: \hspace*{0.5cm} \left( \overline{\underline{3}}, 
 \overline{\underline{2}}, -\frac{1}{6}
   \right) \hspace*{1.5cm} \left( \frac{1}{2}, 0 \right)
\end{equation}
However, the CP-operation recovers a right-handed ``CP-mirror'' with the same 
gauge quantum numbers:
\begin{equation}
 CP \left\{ Q_{L}^{m} \right\} \mapsto Q_{L}^{m c}: \hspace*{0.5cm} 
  \left( \underline{3}, \underline{2}, \frac{1}{6} \right) \hspace*{1.5cm}
   \left( 0, \frac{1}{2} \right)
\end{equation}
When we consider the higgsinos and top quark, it will be useful to 
form 4-component Dirac spinors from the ${\mathcal N}=1$ MSSM fields and the 
CP-conjugates of their mirror fields.  This leads to mixing in the classical
equations of motion as discussed in Appendix 
\ref{app:3}.  Similary for gauginos we associate the usual
 ${\mathcal N}=1$ gaugino and its ${\mathcal N}=2$ superpartner together in a
4-component spinor.  Note that use of the term ``mirror'' will now include 
CP-conjugation implicitly.

\subsection{${\mathcal N}=2$ spinors and 5d kinetic terms} 
\label{app:3}

As discussed in section \ref{sec:ourmodel}, it is convenient to work in terms
of ${\mathcal N}=2$ hypermultiplets.  These are formed from conventional
${\mathcal N}=1$ supermultiplets by adding the CP-conjugates of their 
{\it mirror} superfields with opposite quantum numbers.

  We will consider an explicit example of the third family
that lives in the extra-dimensional bulk.  The third family scalars and their
``mirrors'' are uncoupled, and so only the even parity (MSSM) scalars couple 
directly to the SUSY breaking sector to acquire a soft mass.  This is shown 
in section \ref{sec:spectra}.  However, the form of the 5d Dirac matrices 
causes mixing between fermion fields of even and odd $Z_{2}$-parity.

Consider the top fields charged with respect to the (unbroken)
 $SU(2)_{L}$ gauge group in the MSSM - the left-handed top is contained within
the left-handed quark doublet $q_{3L}$ along with the left-handed bottom 
quark.  The right-handed top is a singlet with 
respect to $SU(2)_{L}$, and so a Dirac mass 
term $\sim m_{t} \left( t_{L} t^{\dagger}_{R} + 
t_{R}^{\dagger} {t}_{L} \right)$ is forbidden by gauge invariance\footnote{A 
Dirac mass may be formed {\it after} the $SU(2)_{L}$ gauge symmetry is broken,
and this is what happens in the (MS)SM through the Higgs mechanism and EWSB.}.

In the ${\mathcal N}=2$ generalisation, we must include additional 
{\it mirror} fields to construct the full 5d hypermultiplet.  The left-handed
top $t_{L}$ and the CP-conjugate of its mirror, $t_{L}^{mc}$, can be combined
into a 4-component Dirac spinor, 
since the charge-conjugated left-handed mirror 
is equivalent to a right-handed fermion.  Similarly for the right-handed 
top $t_{R}$ and its mirror $t_{R}^{m}$.  
Notice that $SU(2)_{L}$ singlets and 
doublets appear in different Dirac spinors, 
and therefore do not break the gauge symmetry.
We have two 4-component Dirac spinors for the top sector, 
where the index labels the handedness of the MSSM fermion:
\begin{equation}
 T_{L} = \left( 
   \begin{array}{c} 
     t_{L} \\
     t_{L}^{mc}
   \end{array} \right) 
\hspace*{2cm}
 T_{R} = \left( 
   \begin{array}{c} 
     t_{R}^{mc} \\
     t_{R}  
   \end{array} \right)
\end{equation}
and similarly ${\displaystyle \bar{T}= T^{\dagger} \gamma^{0} }$
\begin{equation}
 \bar{T}_{L} = \left( 
   \begin{array}{cc} 
     t_{L}^{mc \dagger} & t_{L}^{\dagger}
   \end{array} \right) 
\hspace*{2cm}
 \bar{T}_{R} = \left( 
   \begin{array}{cc} 
     t_{R}^{\dagger} & t_{R}^{mc \dagger}
   \end{array} \right)
\end{equation}

We can now construct the kinetic terms in the 5d lagrangian in the absence
of interactions.
\begin{eqnarray}
   {\mathcal L}_{KE} = i \bar{T}_{L} \gamma^{M} \partial_{M} T_{L} +
          i \bar{T}_{R} \gamma^{M} \partial_{M} T_{R} 
       \hspace*{4.2cm} \nonumber \\
    = i t_{L}^{\dagger} \bar{\sigma}^{\mu} \partial_{\mu} t_{L}
      + i t_{R}^{\dagger} \sigma^{\mu} \partial_{\mu} t_{R}
      + i t_{L}^{mc \dagger} \sigma^{\mu} \partial_{\mu} t_{L}^{mc}
      + i t_{R}^{mc \dagger} \bar{\sigma}^{\mu} \partial_{\mu} t_{R}^{mc} 
         \label{eq:topke} \\
    \hspace*{4cm} 
      - t_{L}^{mc \dagger} \partial_{5} t_{L} 
      + t_{L}^{\dagger} \partial_{5} t_{L}^{mc} 
      + t_{R}^{\dagger} \partial_{5} t_{R}^{mc} 
      - t_{R}^{mc \dagger} \partial_{5} t_{R}   \nonumber 
\end{eqnarray}
Notice that the 4d kinetic terms do not mix fields, while $\gamma^{5}$
leads to mixing between fields and their mirror states.  This leads to 
non-trivial classical equations of motion.

\subsection{Equation-of-motion method details - gauginos} 
\label{app:4}

In this appendix we will show the details of the calculation for the
gaugino mass spectrum using the equation-of-motion method (section 
\ref{sec:m2gauginos}).  We derive Eqn.\ref{eq:gauginos} which is a relation
for the KK-mode mass in terms of the SUSY breaking parameters.
From Eq.\ref{eq:lagrange1}, we see that the even-parity gaugino 
$\lambda_{1}$ couples directly to the SUSY breaking sector at $y= \pi R$
leading to a localized soft gaugino mass.  However, the odd-parity 
${\mathcal N}=2$ 
superpartner $\lambda_{2}$ is coupled to the even gaugino through the 
kinetic term as shown in Appendix \ref{app:3}. 

The gaugino lagrangian includes kinetic terms and the coupling of the 
even-parity gaugino to the SUSY breaking sector, where the even and odd
gauginos form a Dirac spinor
${\displaystyle \Gamma = \left( \lambda_{1} \, \, \lambda_{2} \right)^{T} }$
\begin{eqnarray}
  {\mathcal L}_{\lambda} = i \bar{\Gamma} \gamma^{M} \partial_{M} \Gamma
      + \delta \left( y- \pi R \right) \int d^{2}\theta 
         \frac{c_{w}}{16 g_{5}^{2} M_{\ast}^{2}} 
           S \left( tr W^{\alpha} W_{\alpha} + h.c. \right)
\end{eqnarray}
which leads to the classical equations of motion that couple the two gauginos:
\begin{eqnarray}
  i \sigma^{\mu} \partial_{\mu} \lambda_{2}(x,y) 
   + \partial_{5} \lambda_{1}(x,y) = 0
     \label{eq:gauginomotion1} \\
  i \bar{\sigma}^{\mu} \partial_{\mu} \lambda_{1}(x,y)
   - \partial_{5} \lambda_{2}(x,y)
    - \delta \left( y- \pi R \right) 
      \frac{c_{w}F_{S}}{2 M_{\ast}^{2}} \lambda_{1}(x,y) = 0 
        \label{eq:gauginomotion2}
\end{eqnarray}
Following refs. ~\cite{arkani1,nomura}, we can solve these equation using the 
following solutions:  
\begin{equation}
 \lambda_{i}(x,y) =\sum_{k} \eta_{i,k} (x) g_{i,k} (y)
\end{equation}
where $g_{1,k} \left( g_{2,k} \right)$ are even (odd) with respect to 
the $Z_{2}$-parity transformation $y \rightarrow -y$.  We integrate 
Eqn.\ref{eq:gauginomotion2} in a region $\left( \pi R - \epsilon, \pi R + 
\epsilon
\right)$ where $\epsilon \rightarrow 0$ to obtain boundary conditions at 
$y= \pi R$ which must be satisfied by each KK mode:
\begin{equation}
 \eta_{2,k} (x) = \frac{c_{w} F_{S}}{4 M_{\ast}^{2}} 
  \frac{g_{1,k} \left( \pi R \right)}{g_{2,k} \left( \pi R -\epsilon \right)} 
   \eta_{1,k}(x)  \label{eq:gauginobc}
\end{equation}
Using the 2-component Weyl form of the Dirac equation,
 ${\displaystyle i \bar{\sigma}^{\mu} \partial_{\mu} \eta_{1,k} 
       = m_{\lambda , k} \eta_{1,k} }$ , we can rewrite 
Eqn.\ref{eq:gauginomotion2} at $y \ne \pi R$ 
(so neglecting the delta function)
\begin{eqnarray}
  m_{\lambda,k} \eta_{1,k}(x) g_{1,k}(y) - \frac{c_{w} F_{S}}{4 M_{\ast}^{2}}
   \frac{g_{1,k} \left( \pi R \right)}{g_{2,k} \left( \pi R \right)} 
    \eta_{1,k}(x) \partial_{5} g_{2,k}(y) = 0
\end{eqnarray}
which give solutions\footnote{Notice that this y-dependence is consistent 
with the identification of $\lambda_{1} \left( \lambda_{2} \right)$ with the 
even (odd) parity gaugino states, since we only want $\lambda_{1}$ to couple 
to the boundary fields and have a non-vanishing zero mode which we can 
associate with the MSSM gaugino.}
\begin{equation}
  g_{1,k}(y) \sim \cos \left[ m_{\lambda , k} y \right], \hspace*{1cm}  
      \hspace*{1cm} g_{2,k}(y) \sim \sin \left[ m_{\lambda , k} y \right]
\end{equation}
The final result gives a relation for the gaugino KK-mode masses 
$m_{\lambda,k}$ in terms of the SUSY breaking F-term
\begin{equation}
  \tan \left[ m_{\lambda , k} \pi R \right] = 
   \frac{c_{w} F_{S}}{4 M_{\ast}^{2}}  
\end{equation}


\end{document}